\documentclass[aps,prb,twocolumn,showpacs,superscriptaddress]{revtex4-1}

\usepackage{pdfpages}
\usepackage{relsize}
\usepackage[T1]{fontenc}
\usepackage{amsfonts,amssymb,amsmath}
\usepackage{graphicx,color}
\usepackage{bm,times}
\usepackage[colorlinks=true,citecolor=blue,urlcolor=blue]{hyperref}
\numberwithin{equation}{section}

\makeatletter
\AtBeginDocument{\let\LS@rot\@undefined}
\makeatother

\def\A{\mathrm{A}}

\def\H{\mathrm{H}}

\def\p{\partial}
\def\mb{\mathbf}

\def\AL{\mathrm{AL}}
\def\Im{\mathrm{Im}}
\def\R{\mathrm{R}}
\def\Re{\mathrm{Re}}

\def\pair{\mathrm{pair}}
\def\AL{\mathrm{AL}}
\def\pair{\mathrm{pair}}

\begin{document}

\title{Combined effects of pairing fluctuations and a pseudogap in the Cuprate Hall effect}

\author{Rufus Boyack}
\affiliation{Theoretical Physics Institute and Department of Physics, University of Alberta, Edmonton, Alberta T6G 2E1, Canada}

\author{Xiaoyu Wang}
\affiliation{James Franck Institute, University of Chicago, Chicago, Illinois 60637, USA}

\author{Qijin Chen}
\affiliation{Zhejiang Institute of Modern Physics and Department of Physics, Zhejiang University, Hangzhou, Zhejiang 310027, China}

\author{K. Levin}
\affiliation{James Franck Institute, University of Chicago, Chicago, Illinois 60637, USA}

\begin{abstract}
  The normal-state behavior of the temperature-dependent Hall
  coefficient in cuprate superconductors is investigated using linear response theory.
  The Hall conductivity is of paramount importance
  in that its sign and magnitude directly reflect the sign of the
  charge carriers and the size of particle-hole asymmetry effects.
  Here we apply a strong-pairing fluctuation theory that incorporates
  pseudogap effects known to be important in cuprate transport.
  As a result, in the vicinity of the transition temperature our theoretical approach
  goes beyond the conventional superconducting fluctuation formalism.
  In this regime, pseudogap effects are evident in both
  the transverse and longitudinal conductivities and
  the bosonic response is explicitly gauge invariant.
  The presence of a gap in the excitation spectrum is also apparent at higher
  temperatures, where the gapped fermionic quasiparticles are the
  dominant contribution to the Hall coefficient.
  The observed non-monotonic temperature dependence of the Hall coefficient
  therefore results from a delicate interplay between the fermionic
  quasiparticles and the bosonic fluctuations.
  An important feature of our work is that the sign of the Hall conductivity from the
  Cooper pair fluctuations is the same as that of their fermionic constituents.
  Thus, we find no sign change in the Hall coefficient above the transition temperature.
  This prediction is corroborated by experiments, away from special charge ordering stoichiometries.
  The theoretical results presented in this paper provide crucial
  signatures that can be experimentally verified, enabling validation of the present theory.
\end{abstract}
\maketitle

\section{Introduction}

The behavior of the normal-state Hall coefficient in the copper oxide superconductors continues to be one of the most fundamental characteristics of these materials.
However, like many other normal-state properties, its interpretation is subject to ongoing theoretical debates.
Nevertheless, there is a growing consensus that understanding the Hall coefficient, $R_{\H}$, may help elucidate the origin and nature of the anomalous normal state and the related pseudogap.
Recent attention has focused on ultra-high-field measurements~\cite{Taillefer_2011,Taillefer_2016} where observations of quantum oscillations~\cite{Taillefer_2007} have suggested that pairing effects
may be irrelevant and that fermionic quasiparticles~\cite{Sachdev_2016,Tremblay_2017} are primarily responsible for the behavior of $R_{\H}$.
In contrast to this picture is the extensively discussed~\cite{Fukuyama_1971,Aronov_Rapoport_1992,Aronov_Hikami_Larkin_1995,Michaeli_Tikhonov_Finkelstein_2012,VarlamovBook}
superconducting fluctuation interpretation~\cite{Rice_1991,Samoilov_1994,Lang_Heine_1994,Hwang_1994,Jin_Ott_1998,Konstantinovic_2000,Matthey_2001,Segawa_Ando_2004} of $R_{\H}$ above the transition temperature, $T_{c}$,
where the transport is dominated by bosonic Cooper-pair fluctuations.
The dichotomy between these two approaches is exacerbated by the fact that the conventional fluctuation approach is not directly able to address pseudogap effects which $R_{\H}$ is thought to reflect~\cite{Taillefer_2016}.

The goal of this paper therefore is to incorporate pseudogap effects into a superconducting fluctuation approach to $R_{\H}$.
Even above the transition temperature, $R_{\H}$ is a complicated, non-monotonic function~\cite{Jin_Ott_1998}.
The challenge is to understand both the high-temperature regime in which the Hall coefficient steadily increases with decreasing $T$, as well as the region close to $T_c$, where it rapidly decreases with decreasing $T$.
The theoretical framework implemented in this paper is a strong-pairing fluctuation theory, in contrast to the conventional~\cite{VarlamovBook} weak-pairing fluctuation formalism built upon the Ginzburg-Landau (GL) theory of the BCS regime.
Our theoretical approach provides a qualitative understanding of the temperature and hole-concentration dependence of the low-field cuprate data.

We also address several important aspects involved in the interpretation of cuprate Hall data.
These include debates~\cite{Aronov_Rapoport_1992,Geshkenbein_Ioffe_Larkin_1997} about the sign of the fluctuation contribution to $R_{\H}$, clarifying the significance of the widely observed~\cite{Hwang_1994,Vandermarel_1994,Xiang_2008}
scaling of $R_{\H}$ with the pseudogap onset temperature $T^{*}$, and also identifying future Hall measurements which may elucidate whether the pseudogap persists~\cite{Shibauchi} in high magnetic fields.
Indeed, the pseudogap appears to be inextricably connected to Hall experiments.
There has been particular emphasis attached to the highest hole concentration (called $p^{*}$) at which the pseudogap is non-vanishing~\cite{Taillefer_2016}.
In addition, there are specific hole concentrations at which some form of ordering may take place~\cite{Ando_Segawa_2002,Taillefer_2016}, leading to a possible reconstruction of the Fermi surface and to signatures in $R_{\H}$.

\begin{figure}
\includegraphics[width=0.8\linewidth]{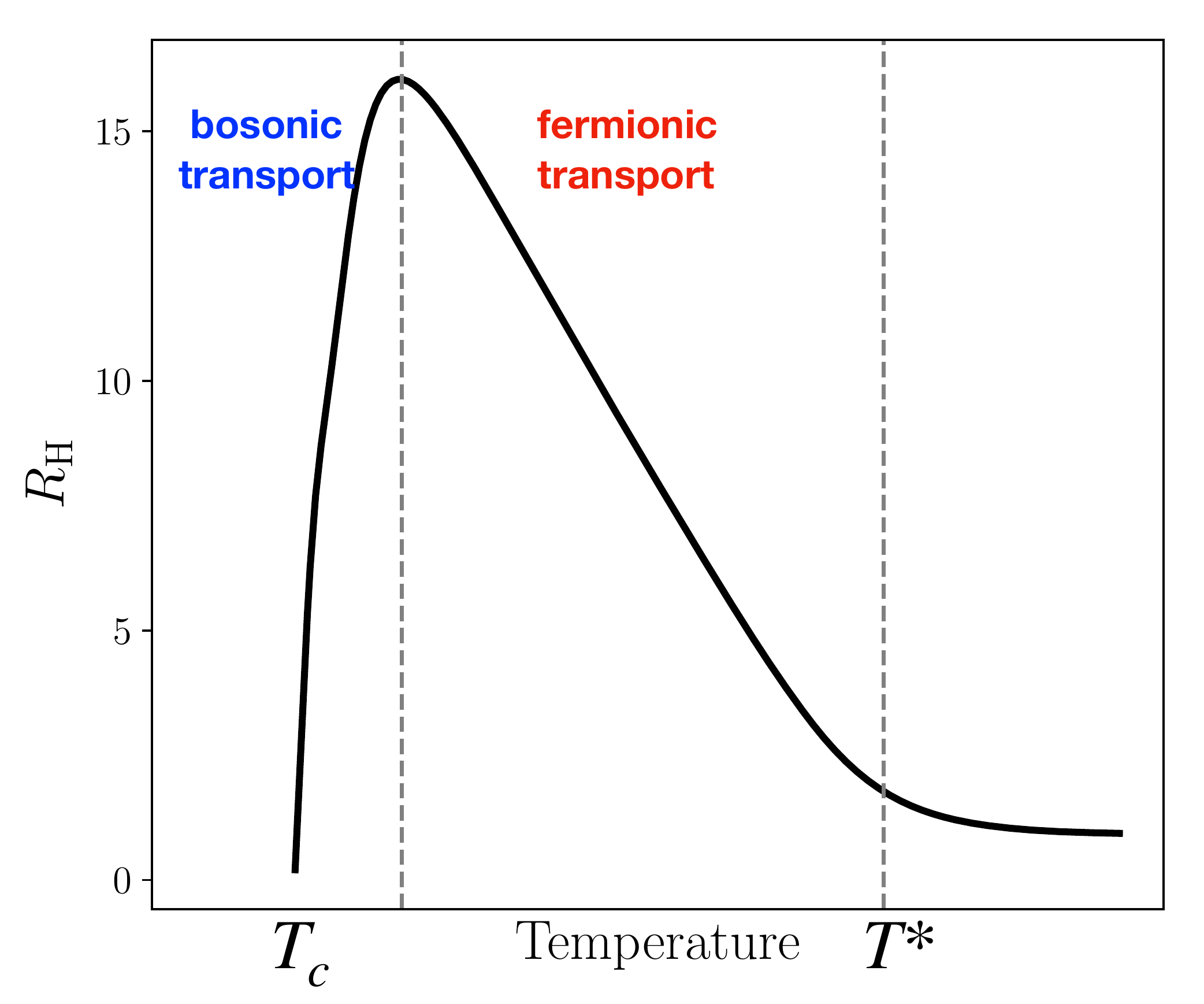}
\caption{Schematic illustration of the non-monotonic temperature dependence of $R_{\H}$.
Near the pseudogap onset temperature, $T^{*}$, the rise of $R_{\H}$ with decreasing $T$ is due to a reduction in the number of unpaired fermions.
Close to $T_{c}$, the rapid decrease in $R_{\H}$ is driven by the current carried by coherent bosonic fluctuations.}
\label{fig:schematic}
\end{figure}

While the literature has generally focused on approaches to $R_{\H}$ that derive from considering only fermionic quasi-particles~\cite{Sachdev_2016,Taillefer_2016,Tremblay_2017,Mitscherling_2018} or only bosonic
fluctuations~\cite{Samoilov_1994,Lang_Heine_1994,Hwang_1994,Jin_Ott_1998,Konstantinovic_2000,Matthey_2001},
in the present strong-fluctuation framework, both bosonic and fermionic degrees of freedom contribute to the Hall coefficient.
Importantly, here the ``bosons'' are composed of gapped fermions, rather than ``free'' fermions as in conventional fluctuation theory.
The fermionic excitation gap, $\Delta(T)$, which reflects the energy needed to break apart the pairs, vanishes at the pseudogap onset temperature $T^{*}$.
In Fig.~\ref{fig:schematic} we present a schematic illustration of $R_{\H}$, showing how its observed~\cite{Rice_1991,Samoilov_1994,Jin_Ott_1998} non-monotonic temperature dependence is driven by an interplay between bosonic and fermionic transport.
The initial rise with decreasing temperature occurs near $T^{*}$ and is due to a reduction in the number of fermionic charge carriers as $\Delta$ increases and fermions convert into bosonic pairs.
The lower temperature region near the superconducting transition $T_{c}$ is dominated by bosonic transport, reflecting the dramatically increasing conductivity due to fluctuating pairs.

The paper is organized as follows.
In Sec.~\ref{sec:Theory}, we begin with a summary of the theoretical literature and then outline the strong-pairing fluctuation approach along with a summary of our transport expressions.
Importantly, here we emphasize the similarities and differences compared to the standard weak-fluctuation approach.
In Secs.~\ref{sec:Boson_Transport}-\ref{sec:Fermion_Transport}, we present detailed derivations of the bosonic and fermionic electrical conductivities.
Our numerical results are then presented in Sec.~\ref{sec:Numerics}. Finally, in Sec.~\ref{sec:Conclusion} we present our conclusions.

\section{Theoretical analysis}
\label{sec:Theory}
\subsection{Previous theoretical treatments}
\label{sec:Previous_Theory}

A theoretical analysis of the Hall conductivity within the context of time-dependent GL theory was first performed by Abrahams et. al~\cite{Abrahams_1971}.
Important to this work was the observation that the transverse electrical conductivity is non-vanishing only when the time-derivative term in the GL equation has an imaginary component.
This GL approach contains only a subset of the contributions which appear in the microscopic superconducting fluctuation formalism~\cite{VarlamovBook}; namely, it includes only the Aslamazov-Larkin (AL) diagram.

Among the earliest attempts to address the Hall conductivity within the microscopic fluctuation formalism were calculations by Fukuyama et. al in Refs.~\cite{Fukuyama_1970,Fukuyama_1971}.
The simple physics behind this diagrammatic analysis is that it contains two contributions to the Hall conductivity:
(i) Maki-Thompson (MT) and Density of States (DOS) terms comprising fermionic scattering mechanisms and (ii) the AL term representing the bosonic Cooper-pair fluctuations.
In Ref.~\cite{Fukuyama_1971}, these authors considered the AL and MT diagrams and showed that the AL contribution is only non-zero provided that one accounts for the energy derivative of the density of states.
Moreover, they also demonstrated that, in contrast to the longitudinal electrical conductivity, where the anomalous MT contribution is of the same order as that of the AL~\cite{VarlamovBook},
the AL contribution to $\sigma_{xy}$ is always more singular than that of the MT.

Subsequent work by Ullah and Dorsey~\cite{Ullah_Dorsey_1991,Dorsey_1992} extended the GL treatment of Hall conductivity by considering the Lawrence-Doniach model of layered superconductors.
The importance of an imaginary time-derivative term in the equations of motion was reiterated in these works and more generally it was recognized that such a term breaks particle-hole symmetry.
In the context of GL theory, general symmetry considerations along with the Onsager relations confirm that a non-vanishing (bosonic) Hall conductivity is obtained only when particle-hole symmetry is broken~\cite{Dorsey_1992,Dorsey_Fisher_1992}.
Dorsey and Fisher~\cite{Dorsey_Fisher_1992} emphasized that the sign of the particle-hole symmetry breaking term is material specific and that the Hall effect thus provides an important probe of the underlying microscopic details of a given superconductor.

After this focus on particle-hole symmetry breaking, a new line of inquiry emerged on the origin and sign of this particle-hole asymmetry term.
This was addressed in more detail in Refs.~\cite{Aronov_Rapoport_1992,Aronov_Hikami_Larkin_1995};
in the latter reference Aronov et. al suggested that from the gauge-invariance of GL theory this term must be proportional to $\partial\ln T_{c}/\partial\ln\mu$.
Aronov and Rapoport~\cite{Aronov_Rapoport_1992} made the argument that under reasonable assumptions the AL Hall conductivity has the same sign as in the normal state.
Moreover, these authors argued that, in general, the AL contribution to Hall conductivity cannot explain the sign change observed near $T_{c}$.

That the sign of the AL Hall conductivity is generally to be associated with the sign of the fermionic Hall conductivity
(determined by the Fermi-surface topology) is an important constraint we emphasize in this paper.
We argue that it derives from the fact that the most natural bosonic degrees of freedom are associated with fermion pairs.
An alternative proposal, however, was developed by Geshkenbein et. al~\cite{Geshkenbein_Ioffe_Larkin_1997}, which arrived at a sign difference between the normal-state fermionic and bosonic Hall conductivities.
In this Bose-Fermi model the particle-hole asymmetry term is not based solely on the energy derivative of the density of states but rather is of fixed magnitude,
and the normal state consists of holes while the bosons are assumed to be formed from electron pairs.

This line of theoretical inquiry has led to little consensus about the sign of the AL contribution to the Hall coefficient in the hole-doped cuprates.
Experiments have established that the generic experimental curves~\cite{Jin_Ott_1998} display a primarily positive and non-monotonic $R_{\H}$ with a possible sign change very close to $T_{c}$.
When this sign change occurs it has been consistently attributed to non-linear field effects~\cite{Ong_1995}.
Interestingly, these same effects are observed in non-cuprate superconductors~\cite{Destraz_2017}.
In general, experiments seem to imply that the sign change may appear either above~\cite{Taillefer_2016,Ando_Segawa_2002} or below~\cite{Jin_Ott_1998}
$T_{c}$ and that the former situation is likely associated with special hole concentrations where there is some degree of charge ordering.
Of paramount importance here, of course, is the definition of $T_c$.
With a magnetic field present, the resistive transition is broad and so the critical temperature cannot be unambiguously established,
in the absence of Meissner data. Although there are alternative choices, it is frequently taken to correspond to the midpoint of the rapidly decreasing longitudinal resistivity (with decreasing $T$) curve~\cite{Comment1}.

The majority of experiments~\cite{Samoilov_1994,Rice_1991,Jin_Ott_1998} have focused on fitting data to the standard expressions~\cite{Fukuyama_1971} for the MT and AL diagrams derived within the conventional fluctuation formalism.
In contrast to the present paper, different signs for these two contributions are generally assumed.
It has been surmised that there is a positive fermionic contribution over the range of temperatures above $T_{c}$ due to the MT term,
while the bosonic contribution from the AL term is presumed to give a large negative contribution just above $T_{c}$.
This latter term, it is argued, would accommodate the possible sign change in the Hall conductivity were it to occur above $T_{c}$.

In addition to interest in the cuprate Hall conductivity, there have also been recent experimental and theoretical studies~\cite{Breznay_2012,Michaeli_Tikhonov_Finkelstein_2012,Varlamov_Galda_Glatz_2018} of the Hall conductivity in disordered thin films.
In Ref.~\cite{Michaeli_Tikhonov_Finkelstein_2012} it was found that,  in addition to the usual ten fluctuation diagrams~\cite{VarlamovBook}, in the presence of a magnetic field there are two additional diagrams.
Another crucial finding in this work, which may bear on the interpretation of cuprate experiments, was that some of the fluctuation diagrams cancel one another,
leaving only the AL, anomalous MT, part of the DOS, and the two new diagrams found, as the remaining contributions to the Hall conductivity.
In the cuprates, the experimental analyses have generally claimed~\cite{Rice_1991} that for temperatures far greater than $T_{c}$ it is the MT contribution that is significant.
Nonetheless, it is  important to ensure that whatever experimental fitting procedure is adopted, it must be reconciled with this more recent theoretical work~\cite{Michaeli_Tikhonov_Finkelstein_2012}.

The AL and DOS class of diagrams represent the bosonic and fermionic contributions, respectively, and they provide the basis for the Hall conductivity calculations performed in the present paper.
Our formalism is based on a strong-pairing fluctuation theory~\cite{Boyack_2018} that goes beyond the weak-pairing formalism~\cite{VarlamovBook},  which omits the important normal-state gap.
As will be discussed in the next section, this strong-pairing theory naturally incorporates a particle-hole asymmetric term with a sign determined by the sign of the fermionic (hole-like) charge carriers.
The approach we use is not limited to a small temperature scale $\sim T_{c}/E_{F}$ as in the conventional framework~\cite{VarlamovBook}.
Since we address a larger range of temperatures this means the effects from the fermionic quasiparticle excitation gap must necessarily be included.
It is also important to emphasize that in this strong-pairing fluctuation approach the Cooper pairs are more stable than in weak-fluctuation theory.
This is a consequence of the excitation (pseudo)gap which the fermions experience; it is this gap which impedes their decomposition from composite Cooper pairs into individual fermions.

The next section gives an overview of our theoretical framework and a summary of our electrical conductivity results.

\subsection{Overview of the strong-pairing fluctuation theory}

It is useful to now present a more detailed summary of the strong-pairing fluctuation theory on which this paper is based.
This approach presumes that a stronger-than-BCS attraction is present and belongs to a class of BCS--Bose Einstein condensation (BEC) crossover theories.
What distinguishes it from others in this class~\cite{Nozieres_SchmittRink_1985} is that it is founded upon an equation of motion approach with a systematic Green's function decoupling scheme~\cite{Kadanoff_Martin_1961},
which was shown to be consistent with the underlying structure of BCS theory.
The extension to the case where the interaction strength is arbitrary~\cite{Chen_Stajic_2005} relates to the BCS-Leggett~\cite{Leggett_1980} ground state,
and within this generalization of BCS theory we are able to address finite temperature effects \cite{Chen_Stajic_2005}.
In the theoretical analysis we use natural units: $c=\hbar=k_{B}=1$; these units are restored when appropriate.

In terms of the small four-vector $q^{\mu} = (\Omega, \mb{q})$, the inverse (retarded) pair-propagator can be generically written as
\begin{equation}
\label{eq:Tmat_propagator}
t^{-1}(q)\approx Z[\kappa\Omega-\mb{q}^{2}/\left(2M_{\pair}\right)-|\mu_{\pair}|+i\Gamma\Omega].
\end{equation}
The coefficients $\kappa$ and $\Gamma$ are real and dimensionless.
The real part of $t^{-1}(q)$ contains contributions which depend on an effective pair mass, $M_{\pair}$, and a pair chemical potential $\mu_{\pair}=t^{-1}(0)/Z$.
Except in the particle-hole symmetric case (where $\kappa=0$ and an $\Omega^2$ term would be included) in general we have $|\kappa| =1$.
The sign of $\kappa$ indicates whether the pairs consist of pairs of electrons $(+1)$ or pairs of holes $(-1)$, as explained below.
Additionally, the imaginary part, $\propto\Gamma\Omega$, represents the diffusive contribution to the inverse pair propagator.
In the actual numerical calculations an anisotropic pair dispersion is used:
$\Omega_\mb{q} = q_\parallel^2/(2M_\parallel) + q_\perp^2/(2M_\perp)$, reflecting the layered structure of the cuprates,
where ``$\parallel$'' and ``$\perp$'' denote in-plane and out-of-plane hopping, respectively.

As expected in a strong-fluctuation theory, the pseudogap, $\Delta$, must appear in the pair propagator.
This is in contrast to the weak-fluctuation theory where the propagator consists of only bare fermions.
The incorporation of the pseudogap arises through the dressing of a single Green's function via:
\begin{equation}
\label{eq:Tmat}
t^{-1}(q)\equiv g^{-1}+\sum_{k}G(k)G_{0}(-k+q)\varphi_{\mb{k}-\mb{q}/2}^2.
\end{equation}
Here, $q^{\mu}=(i\Omega_{m},\mb{q})$, $k^{\mu}=(i\omega_{n},\mb{k})$ (before analytic continuation) where $\Omega_{m}$ and $\omega_{n}$ are bosonic and fermionic Matsubara frequencies, respectively.
The $d$-wave form factor is $\varphi_{\mb{k}}$. The summation is defined as $\sum_{k}\equiv T\sum_{i\omega_{n}}\int d^dk/(2\pi)^d$.
The dressed and bare electron Green's functions are $G$ and $G_{0}$, respectively.
That only one dressed Green's function appears in the pair propagator has been extensively discussed in the literature~\cite{Maly_1999,Chen_Stajic_2005,Boyack_2018}
and is understood to be a direct consequence of an equation of motion approach to generalizing BCS theory~\cite{Kadanoff_Martin_1961}.

We emphasize that the pair propagator can assume either an electron-like or hole-like character depending on the constituent fermions
(through the Fermi surface curvature determined by the band parameters) and this is associated with a sign change in $\kappa =\pm 1$.
As a corollary, hole-like quasiparticles (with positive $R_{\H}$) lead to hole-like Cooper pairs and electron-like quasiparticles (with negative $R_{\H}$) are associated with electron-like pairs.
In the hole-like case, the normal-state $\sigma_{xy}$ is positive, while $\kappa$ is negative and as a result the Cooper-pair $\sigma_{xy}$ is also positive [see Eq.~(\ref{eq:sigmaxy_AL3d})].

The Hall coefficient, $R_{\H}$, is defined by
\begin{equation}
R_{\H}=\frac{E^{y}}{J^{x}B^{z}},
\end{equation}
where $E^{y}, J^{x}$, and $B^{z}$ are the corresponding components of the electric field, current, and magnetic field, respectively.
The linear constitutive relations between $\mb{E}$ and $\mb{J}$ are $\mb{E}=\tensor{\rho}\mb{J}$ and, equivalently, $\mb{J}=\tensor{\sigma}\mb{E}$, with $\tensor{\rho}$ and $\tensor{\sigma}$ the resistivity and conductivity tensors, respectively.
In the absence of a $y$-component to the current $J^{y}=0$ and the first constitutive relation gives $E^{y}=\rho_{yx}J^{x}$, and thus $\rho_{yx} = B R_{\H}$.
In terms of the conductivity tensor elements this becomes
\begin{equation}
\label{eq:Hall_RH}
R_{\H}=\frac{1}{B}\frac{\sigma_{xy}}{\sigma_{xx}^{2}+\sigma_{xy}^{2}}.
\end{equation}
Here we have used the fact that $\sigma_{xx}=\sigma_{yy}$ and $\sigma_{xy}=-\sigma_{yx}$.
Throughout the paper, the fermionic ($f$) and bosonic ($b$) contributions are added together,
and when computing $R_{\H}$ their sum enters directly in the conductivities via: $\sigma_{ij}=\sigma_{ij}^{f}+\sigma_{ij}^{b}$.

To make contact with the conventional fluctuation literature, note that the coherence length which generally appears~\cite{VarlamovBook} is now replaced by
\begin{equation}
\xi(T)\equiv\xi_{0}/\sqrt{\epsilon}\rightarrow\xi_{0}/\sqrt{|\mu_{\pair}|/(k_{B}T_{c})}
\end{equation}
In the conventional fluctuation literature, $\epsilon\equiv\ln(T/T_{c})\approx(T-T_{c})/T_{c}$, is the reduced temperature and corresponds to the dimensionless parameter characterizing the transport singularities near $T_{c}$;
in the strong-pairing theory this is replaced by the rescaled bosonic pair chemical potential $\epsilon \rightarrow |\mu_{\pair}|/(k_{B}T_{c})$.
Similarly the (zero-temperature) coherence length now becomes~\cite{Comment3}
\begin{equation}
\xi_{0}=\hbar\sqrt{\frac{1/(2M_{\pair})}{k_{B}T_{c}}}.
\end{equation}
For the bosonic contribution we may anticipate the answers for $R_{\H}$ (obtained from detailed microscopic analysis) by using the above
correspondences along with previous~\cite{Fukuyama_1971,VarlamovBook,Varlamov_Galda_Glatz_2018} fluctuation calculations.
In the conventional fluctuation theory the bosonic Hall conductivity is proportional to the particle-hole asymmetry term in the GL propagator~\cite{VarlamovBook}, which is dependent on Fermi-surface topology~\cite{Angilella_2003},
and in the strong-fluctuation theory this corresponds to $\kappa$, whose sign is determined by the nature of the Fermi-surface.
The non-singular fermionic contribution can be similarly anticipated from the usual quasiparticle approximation~\cite{Tremblay_2017} to the Hall effect.
It appears primarily as a density of states term which now includes the pseudogap in the fermionic dispersion.

A microscopic analysis of the bosonic transport coefficients is presented in Sec.~\ref{sec:Boson_Transport}.
Here we summarize the results for the bosonic contributions to the two-dimensional (2D) electrical conductivity tensor:
\begin{align}
\label{eq:sigmaxx_Boson}
\sigma_{xx}^{b} &= \frac{\left(e^{*}\right)^{2}}{\hbar}\frac{(\kappa^{2}+\Gamma^{2})}{\Gamma}\frac{k_{B}T}{8\pi|\mu_{\pair}|},\\
\label{eq:sigmaxy_Boson}
\frac{\sigma_{xy}^{b}}{B} &= -\frac{\left(e^{*}\right)^{2}}{\hbar}\frac{\hbar e^{*}}{M_{\pair}c}\frac{\kappa(\kappa^{2}+\Gamma^{2})}{\Gamma^{2}}\frac{k_{B}T}{48\pi|\mu_{\pair}|^{2}}.
\end{align}
The charge of the bosonic fluctuations is $e^{*}=2e$; in this paper we adopt the convention $e>0$ so that the charge of the electron is $q_{e}=-e$.
The constants $\hbar, c,$ and $k_{B}$ have been restored here.
The Gaussian units of 2D conductivity (actually a conductance) are those of $e^2/\hbar$, and this is explicit in the above expressions.

The combination $\hbar e^{*}B/(M_{\pair}c)$ is equivalent to the energy $\hbar\omega^{*}_{c}$, which depends on the bosonic cyclotron frequency $\omega^{*}_{c}=e^{*}B/(M_{\pair}c)$.
The general structure of a linear response treatment of fluctuation theory in the presence of a magnetic field requires that $B$ appears in bosonic transport coefficients in a perturbative fashion in powers of a small dimensionless parameter:
\begin{equation}
\label{eq:EnergyScale}
\left|\frac{\hbar\omega^{*}_{c}}{\mu_{\pair}}\right| \ll 1.
\end{equation}

The fermionic conductivity tensor is derived in Sec.~\ref{sec:Fermion_Transport}. Presuming positive charge carriers, this is given by
\begin{align}
\label{eq:sigmaxx_Fermion}
\sigma_{xx}^{f} &= 2e^{2}\tau\sum_{\mb{k}}v^{2}_{x}\left(\frac{\xi_{\mb{k}}}{E_{\mb{k}}}\right)^{2}\left(-\frac{\p f(E_{\mb{k}})}{\p E_{\mb{k}}}\right),\\
\label{eq:sigmaxy_Fermion}
\frac{\sigma_{xy}^{f}}{B} &= \frac{e^{3}\tau^{2}}{2c}\sum_{\mb{k}}\left(v_{x}^{2}v_{yy}-v_{x}v_{y}v_{xy}\right)\left(1+\frac{3\xi_{\mb{k}}^{2}}{E_{\mb{k}}^{2}}\right)\nonumber \\
& \quad\times\left(-\frac{\partial f(E_{\mb{k}})}{\partial E_{\mb{k}}}\right).
\end{align}
Here, $\tau$ is a phenomenological parameter representing the quasiparticle lifetime. The quasiparticle dispersion $E_{\mb{k}}$ obeys
$E_{\mb{k}}^{2}=\xi_{\mb{k}}^{2}+\left|\Delta\varphi_\mathbf{k}\right|^{2}$, where $\xi_{\mb{k}}$ is the bandstructure, and the velocities are $v_{i}=\partial\xi_{\mb{k}}/\partial k^{i}, v_{ij}=\partial^{2}\xi_{\mb{k}}/\partial k^{i}\partial k^{j}$.
The Fermi-Dirac distribution function is $f(x)=\left(\exp(\beta x)+1\right)^{-1}$.

The focus of this paper is the weak magnetic field regime where the $y$-$x$ component of the resistivity is linear in the magnetic field: $\rho_{yx} = B R_{\H}$ and $R_{\H}$ is field independent.
This is consistent with magnetic fields up to a few Tesla~\cite{Iye_1989}. From Eq.~(\ref{eq:Hall_RH}), this implies that $R_{\H}$ can be approximated as
\begin{equation}
\label{eq:rhoyx}
R_{\H}\approx \frac{1}{B}\frac{\sigma_{xy}}{\sigma^{2}_{xx}}.
\end{equation}
Thus, $\sigma_{xy}^2$ in the denominator of Eq.~(\ref{eq:Hall_RH}) can be dropped, which implies that  $|\sigma_{xy}|\ll|\sigma_{xx}|$.
Indeed, experimentally it is found~\cite{Jin_Ott_1998} that, even in moderately large magnetic fields and for general temperatures, $|\rho_{yx}/\rho_{xx}|\approx 10^{-2}$,
or equivalently, $|\sigma_{xy}/\sigma_{xx}|\approx10^{-2}$.
The criterion in Eq.~(\ref{eq:rhoyx}) then defines precisely what is meant by the weak magnetic field regime.
Using Eqs.~(\ref{eq:sigmaxx_Boson}-\ref{eq:sigmaxy_Boson}), the ratio of the 2D Hall conductivity to the 2D longitudinal conductivity is
\begin{equation}
\sigma_{xy}/\sigma_{xx}=-\frac{\kappa}{6\Gamma}\frac{\hbar\omega^{*}_{c}}{|\mu_{\pair}|}.
\end{equation}
Thus, from an experimental perspective, we again arrive at the constraint in Eq.~(\ref{eq:EnergyScale}).
Note also, by limiting our focus in this paper to the weak magnetic field regime, we do not consider a possible near-$T_{c}$ sign change~\cite{Jin_Ott_1998} in $R_{\H}$,
which has been shown~\cite{Ong_1995,Destraz_2017} to be associated with non-linear field effects and may be relevant only below the zero-field transition temperature $T_{c0}$~\cite{Segawa_Ando_2004,Ando_2004}.

After inserting the 2D conductivity expressions from Eqs.~(\ref{eq:sigmaxx_Boson}-\ref{eq:sigmaxy_Boson}) into Eq.~(\ref{eq:rhoyx}), the more strongly temperature dependent feature
associated with $\mu_{\pair}$ cancels out from the ratio and $R_{\H}$ is predicted to asymptote to a more weakly temperature dependent functional form, as $T$ is decreased.
This result appears at odds with experiments, where the Hall coefficient rapidly decreases as the superconducting transition is approached~\cite{Jin_Ott_1998}; this is evident even away from the immediate vicinity of $T_{c}$.
(This cancellation of $\mu_{\pair}$ is also found for the conventional fluctuation theory, with $\mu_{\pair}$ replaced by $\epsilon$.)
The data provides an important clue~\cite{Jin_Ott_1998,Ando_Segawa_2002}:
the plummeting of $\rho_{yx}$ must be associated with the temperature dependence of the denominator in Eq.~(\ref{eq:rhoyx}) as there is relatively less $T$ variation in $\sigma_{xy}$.

A reconciliation of the experimentally-measured Hall coefficient within conventional fluctuation theory was presented in Ref.~\cite{Breznay_2012} in a systematic study of disordered thin films of the superconductor TaN.
These authors noted that as the transition was approached the AL expression for $\sigma_{xx}$ no longer fit the data and that the divergence in the longitudinal conductivity was stronger than expected.
To address this issue the authors appealed to inhomogeneity effects~\cite{Char_Kapitulnik_1988}, where the predicted dependence in the 2D longitudinal conductivity changes from $\sigma_{xx}^b \propto \frac{1}{\epsilon}$ to
\begin{equation}
\label{eq:sigmaxx_alpha}
\sigma_{xx}^b \propto \frac{1}{\epsilon^{1 + \alpha}},
\end{equation}
where $\alpha \approx 1/3$~\cite{Char_Kapitulnik_1988}.
At the same time they suggested that $\sigma_{xy}$ is unaffected by inhomogeneity effects~\cite{Breznay_2012}.
With this modification it was demonstrated that superconducting fluctuation theory (albeit in a conventional superconductor) can successfully address Hall data.

This provides the motivation for a (single) additional assumption in the current paper.
Indeed, while the cuprates are thought to be clean in the sense of having negligible contamination from impurities, they have been shown to have intrinsic disorder~\cite{Yazdani_2007}.
Along with bulk disorder signatures~\cite{Niedermayer_1998,Panagopoulos_2002},  (surface) scanning tunneling microscopy has led investigators~\cite{Davis_2007} to characterize the cuprates as ``electronic glasses".
For this reason, we incorporate this phenomenological assumption and adopt Eq.~(\ref{eq:sigmaxx_alpha}).
With this inclusion the otherwise microscopic equations (as written above) provide a rather complete qualitative picture of the behavior of the Hall coefficient in the cuprates, which is summarized in Fig.~\ref{fig:schematic}.
In the numerical section (\ref{sec:Numerics}) of the paper we will discuss specific features and quantitative plots of Hall response after first presenting the theoretical formalism.

\section{Bosonic electromagnetic response}
\label{sec:Boson_Transport}
\subsection{Bosonic longitudinal conductivity}
\label{sec:Boson_LongCond}

We begin with the longitudinal fluctuation conductivity in zero external magnetic field, noting that for the strong-pairing fluctuation theory studied in this paper the exact EM response has previously been determined~\cite{Boyack_2018}.
For small wavevectors, the pair propagator corresponds to a quadratically dispersing boson with charge $e^{*}=2e$, renormalized mass $M_{\pair}$, and chemical potential $\mu_{\pair}$.
The exact AL diagram can then be viewed as the response of an effectively free boson, but importantly with vertices constructed from the propagator in a self-consistent manner; see Ref.~\cite{Boyack_2018} for details.

\begin{figure}[h]
\includegraphics[width=.6\linewidth,clip]{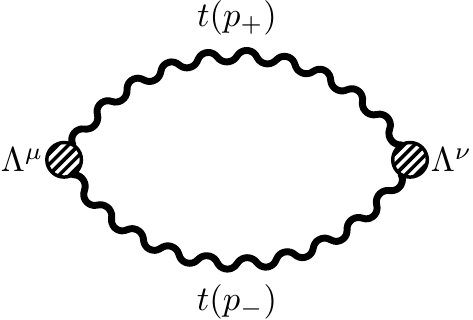}
\caption{Feynman diagram for the bosonic two-particle function.}
\label{fig:AL_xx}
\end{figure}

The longitudinal component of the electrical conductivity is computed using the standard Kubo formula~\cite{MahanBook}:
\begin{equation}
\label{eq:sigmaxx}
\sigma_{xx}=-\underset{\Omega,\mb{q}\rightarrow0}{\mathrm{lim}}\frac{1}{\Omega}\mathrm{Im}\left[\left.P^{xx}(\Omega,\mb{q})\right|_{i\Omega_{m}\rightarrow\Omega+i0^{+}}\right].
\end{equation}
The bosonic two-particle correlation function is
\begin{equation}
\label{eq:Boson_Response}
P_{\AL}^{xx}(q)=-\left(e^{*}\right)^{2}\sum_{p}t(p_{+})\Lambda^{x}(p_{+},p_{-})t(p_{-})\Lambda^{x}(p_{-},p_{+}).
\end{equation}
The bosonic four-vector $p^{\mu}=(i\varpi_{m},\mb{p})$, where $\varpi_{m}$ is a bosonic Matsubara frequency, and $p_{\pm}^{\mu}\equiv p^{\mu}\pm q^{\mu}/2$.
The four-vector summation is defined by $\sum_{p}=T\sum_{i\varpi_{m}}\int\mathrm{d}^dp/(2\pi)^d$.
The EM vertices are bosonic vertices constructed such that $q_{\mu}\Lambda^{\mu}(p_{+},p_{-})=t^{-1}(p_{+})-t^{-1}(p_{-})$.
A bosonic equivalent of the Ward identity~\cite{Ryder} between the vertex and the propagator can be obtained by taking the limit $q\rightarrow0$, which results in $\Lambda^{\mu}(p,p)=\partial t^{-1}(p)/\partial p_{\mu}$.
This important constraint between the fluctuation propagator and the bosonic vertex shows that they must be treated on an equal footing.
In the strong-pairing fluctuation theory these vertices have been determined exactly~\cite{Boyack_2018}.
For the special case where the pair-propagator takes the form given in Eq.~(\ref{eq:Tmat_propagator}), the more complicated triangular vertices of the AL diagram reduce simply to
$\Lambda^{\mu}(p_{+},p_{-})=Z(\kappa+i\Gamma,\mb{p}/M_{\pair})$. The diagram for the bosonic two-particle response is shown in Fig.~\ref{fig:AL_xx}.

The next step is to calculate the Matsubara frequency summation~\cite{VarlamovBook} in Eq.~(\ref{eq:Boson_Response}),
then perform analytic continuation to real frequencies: $i\Omega_{m}\rightarrow\Omega+i0^{+}$, and finally take the limit $\mb{q}\rightarrow0$.
This procedure results in
\begin{align}
\label{eq:AL_xx}
P_{\AL}^{xx}(\Omega,0) &= -\left(e^{*}\right)^{2}\sum_{\mb{p}}\left(\frac{Zp^{x}}{M_{\pair}}\right)^{2}\int_{-\infty}^{\infty}\frac{d\varpi}{2\pi}\coth\left(\frac{\beta\varpi}{2}\right)\nonumber \\
 &\quad \times\Im\ t_{\R}(\varpi,\mb{p})\left[t_{\R}(\varpi+\Omega,\mb{p})+t_{\A}(\varpi-\Omega,\mb{p})\right].
\end{align}
Here, $t_{\R}$ is the retarded pair propagator, as appears in Eq.~(\ref{eq:Tmat_propagator}), and $t_{\A}$, the advanced pair-propagator, corresponds to the complex conjugate of $t_{\R}$.
After taking the limit $\Omega\rightarrow0$ in Eq.~(\ref{eq:AL_xx}) and then integrating by parts, the AL contribution to the longitudinal electrical conductivity is
\begin{equation}
\sigma_{xx}^{b}=\frac{\left(e^{*}\right)^{2}}{2T}\sum_{\mb{p}}\left(\frac{Zp^{x}}{M_{\pair}}\right)^{2}\int_{-\infty}^{\infty}\frac{d\varpi}{2\pi}\frac{\left[\Im\ t_{\R}(\varpi,\mb{p})\right]^{2}}{\sinh^{2}\left(\beta \varpi/2\right)}.
\end{equation}
In the small $|\mu_{\pair}|$ limit ($|\mu_{\pair}|/T_{c}\ll1$) the dominant contribution to the integral occurs when $\beta\varpi\ll1$, which allows the $\sinh$ function to be expanded as $\sinh(\beta \varpi/2)\approx \varpi/(2T)$.
After inserting the pair-propagator from Eq.~(\ref{eq:Tmat_propagator}), then computing the frequency integration, followed by the momentum
integration, we obtain Eq.~(\ref{eq:sigmaxx_Boson}) for $d=2$ and for $d=3$ the result is
\begin{equation}
\sigma_{xx}^{b}=\frac{(\kappa^{2}+\Gamma^{2})}{\Gamma}\frac{k_{B}T\left(e^{*}\right)^{2}}{8\pi\hbar^{2}}\sqrt{\frac{M_{\pair}}{2|\mu_{\pair}|}}.
\end{equation}
The constants $\hbar$ and $k_{B}$ have been restored; the Gaussian units of (3D) conductivity are $s^{-1}$, which are those of the above expression.
Note that $\sigma_{xx}^{b}$ is independent of the signs of both $e$ and $\kappa$.
The longitudinal electrical conductivity is thus the same for electrons and holes and furthermore it is independent of the sign of the particle-hole asymmetry term $\kappa$.

\subsection{Bosonic transverse conductivity}
\label{sec:Boson_TranvCond}

To determine the fluctuation contribution to the transverse, magnetic field dependent conductivity $\sigma_{xy}$, a three-particle EM response must be computed.
While in the weak-pairing fluctuation theory electromagnetic transport is often derived~\cite{VarlamovBook} from a fluctuation free energy, this is not possible in the strong-pairing formalism since it is not phi-derivable.
In principle, one could perform all EM vertex insertions in the two-particle EM response, however, such an approach is intractable.
To make progress, we build upon the analysis of the previous section and consider the response of quasi-independent bosons described by the propagator in Eq.~(\ref{eq:Tmat_propagator}).

The physical situation under consideration consists of measuring the current in the $\hat{x}$-direction in response to applied electric and magnetic fields in the $\hat{y}$ and $\hat{z}$-directions, respectively.
The magnetic vector potential is thus $\mb{A}=c/(i\Omega)\mb{E}e^{-i\Omega t}+B/(iQ)\hat{x}e^{-iQ\hat{y}\cdot\mb{r}}$.
For generality, in the analysis below the generic three-particle EM response function $K^{\mu\nu\alpha}(i\Omega_{m},\mb{q})$ is studied.
An important point not widely appreciated in the fluctuation literature is the need to incorporate two classes of correlation functions, namely, current-current-current $(K_{JJJ})$ and current-density $(K_{J\rho})$.
This necessity is required in order to obtain a gauge invariant three-particle EM response.
Indeed, Fukuyama et. al~\cite{Fukuyama1_1969,Fukuyama2_1969} proved that both $K_{JJJ}$ and $K_{J\rho}$ correlation functions must be included in the full three-particle EM response to render it gauge-invariant.
A detailed discussion of gauge invariance for the bosonic response is deferred to Appendix~\ref{sec:Gauge_Invariance}.

\begin{figure}[h]
\includegraphics[width=\linewidth,clip]{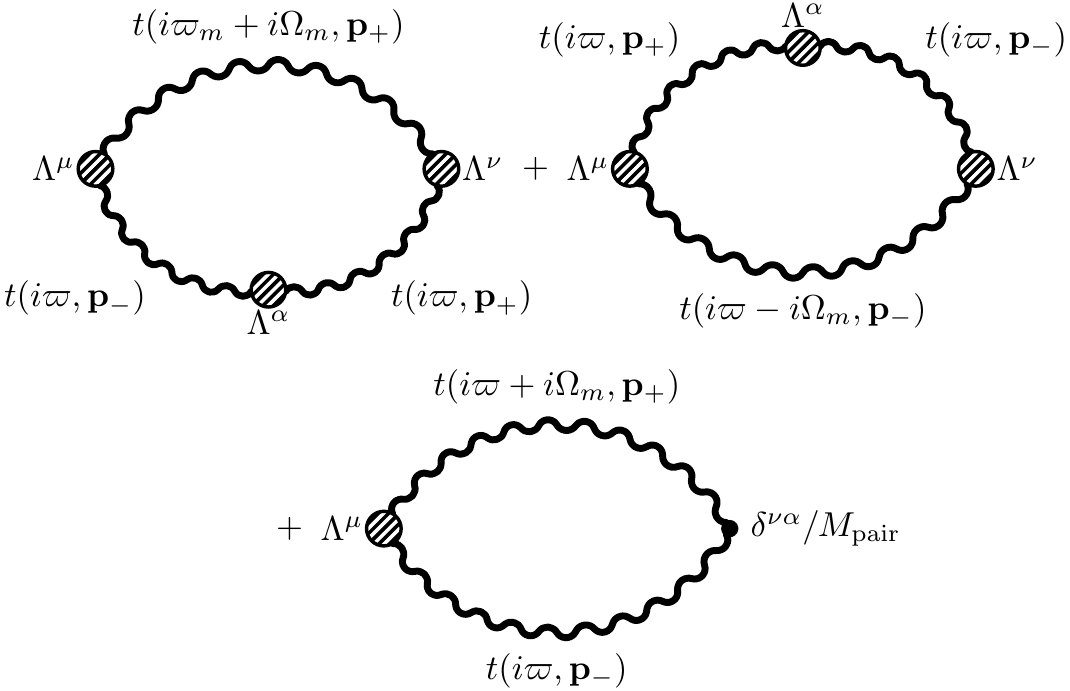}
\caption{Feynman diagrams for the bosonic three-particle function.}
\label{fig:AL_xy}
\end{figure}

The current-current-current and current-density correlation functions are given by~\cite{Fukuyama1_1969,Voruganti_1992}:
\begin{widetext}
\begin{align}
\label{eq:KJJJ}
K_{JJJ}^{\mu\nu\alpha}(q) &= \left(Ze^{*}\right)^{3}\sum_{p}\biggl[\frac{\mb{p}_{+}^{\nu}}{M_{\pair}}\left(\frac{\mb{p}^{\mu}}{M_{\pair}}\frac{\mb{p}^{\alpha}}{M_{\pair}}\right)t(i\varpi_{m}+i\Omega_{m},\mb{p}_{+})t(i\varpi_{m},\mb{p}_{-})t(i\varpi_{m},\mb{p}_{+})\nonumber \\
 & \hspace{1.55cm}+\frac{\mb{p}_{-}^{\nu}}{M_{\pair}}\left(\frac{\mb{p}^{\mu}}{M_{\pair}}\frac{\mb{p}^{\alpha}}{M_{\pair}}\right)t(i\varpi_{m}-i\Omega_{m},\mb{p}_{-})t(i\varpi_{m},\mb{p}_{+})t(i\varpi_{m},\mb{p}_{-})\biggr],\\
\label{eq:Krho}
K_{J\rho}^{\mu\nu\alpha}(q) &= \frac{\left(Ze^{*}\right)^{3}}{M_{\pair}}\delta^{\nu\alpha}\sum_{p}\left(\frac{\mb{p}^{\mu}}{M_{\pair}}\right)t(i\varpi_{m}+i\Omega_{m},\mb{p}_{+})t(i\varpi_{m},\mb{p}_{-}).
\end{align}
\end{widetext}
The current vertex indices are denoted by $\mu,\nu,\alpha\in\{x,y\}$ and $\mb{p}_{\pm} = \mb{p}\pm\mb{q}/2$, where for the case of physical interest $\mb{q}=Q\hat{\mb{y}}$.
The diagram for the general three-particle bosonic response is shown in Fig.~\ref{fig:AL_xy}.
The details of the Matsubara frequency summation are presented in Appendix~\ref{sec:ALHall_Derivation} and here we just quote the final result.
After  taking the limit $\mb{q}\rightarrow0$, followed by $\Omega\rightarrow0$, the real part of the three-particle EM response reduces to
\begin{widetext}
\begin{equation}
\label{eq:AL_xy}
\mathrm{Re}K^{\mu\nu\alpha}\left(q\right)=\left(\mb{q}^{\nu}\delta^{\mu\alpha}-\mb{q}^{\mu}\delta^{\nu\alpha}\right)\frac{\Omega\left(Ze^{*}\right)^{3}}{dM_{\text{pair}}}\sum_{\mb{p}}\left(\frac{p}{M_{\text{pair}}}\right)^{2}
\int_{-\infty}^{\infty}\frac{d\varpi}{2\pi}\coth\left(\frac{\beta\varpi}{2}\right)\left[\Re(\p_{\varpi}t_{\R})\Im(t_{\R}^{2})-\Re(\p_{\varpi}t_{\R}^{2})\Im(t_{\R})\right].
\end{equation}
\end{widetext}
The spatial dimensionality is $d$ and $t_{\R}\equiv t_{\R}(\varpi,\mb{p})$.
The first term arises from the $JJJ$ correlation function, whereas the second term arises from the $J\rho$ contribution.
Due to the prefactor $\left(\mb{q}^{\nu}\delta^{\mu\alpha}-\mb{q}^{\mu}\delta^{\nu\alpha}\right)$ appearing here,
this expression is manifestly gauge-invariant ~\cite{Fukuyama1_1969,Fukuyama2_1969}: $\mb{q}_{\alpha}\mathrm{Re}K^{\mu\nu\alpha}\left(q\right)=0$.
It is important to reiterate that satisfying gauge invariance requires the current-density correlation function to be included, as shown for arbitrary momentum in Appendix~\ref{sec:Gauge_Invariance}.

Finally, the transverse conductivity can be computed using the Kubo formula~\cite{Fukuyama1_1969,Voruganti_1992}:
\begin{equation}
\label{eq:sigmaxy_coeff}
\frac{\sigma_{xy}^{b}}{B}=\underset{\Omega,Q\rightarrow0}{\mathrm{lim}}\frac{1}{\Omega Qc}\mathrm{Re}\left[\left.K^{xyx}(q)\right|_{i\Omega_{m}\rightarrow\Omega+i0^{+}}\right].
\end{equation}
Applying this definition to the correlation function in Eq.~(\ref{eq:AL_xy}) we find that only the $K_{JJJ}$ term contributes, while the $K_{J\rho}$ term vanishes.
An equivalent definition for $\sigma_{yx}$ can be given using the above formula but with $K^{yxx}$ used instead.
In this case the $K_{JJJ}$ term vanishes and it is $K_{J\rho}$ that contributes.
However, since $\sigma_{yx}=-\sigma_{xy}$ and $K^{yxx}=-K^{xyx}$, both expressions above give exactly the same Hall conductivity.
This proves that previous (non-gauge-invariant) calculations of the AL contribution to $\sigma_{xy}$, based solely on the $K_{JJJ}$ term, are unaltered by the inclusion of the $K_{J\rho}$ correlation function.
This is an explicit consequence of the gauge-invariance of the three-particle EM response.
Note, however, for an arbitrary anisotropic dispersion both three-particle response functions contribute; see Refs.~\cite{Voruganti_1992,Tremblay_2017,Mitscherling_2018}

Inserting the response function from Eq.~(\ref{eq:AL_xy}) into Eq.~(\ref{eq:sigmaxy_coeff}), and then integrating by parts, results in
\begin{equation}
\label{eq:sigmaxy_BosonKubo}
\frac{\sigma_{xy}^{b}}{B}=\frac{\beta\left(Ze^{*}\right)^{3}}{3M_{\pair}c}\sum_{\mb{p}}\left(\frac{p^{x}}{M_{\pair}}\right)^{2}\int_{-\infty}^{\infty}\frac{d\varpi}{2\pi}\frac{\left[\Im\ t_{\R}(\varpi,\mb{p})\right]^{3}}{\sinh^{2}\left(\beta\varpi/2\right)}.
\end{equation}
For a particle-hole symmetric fluctuation propagator, where
$\mathrm{Im}\left[t_{\R}\left(-\Omega,\mb{p}\right)\right]=-\mathrm{Im}\left[t_{\R}\left(\Omega,\mb{p}\right)\right]$ and
$\mathrm{Re}\left[t_{\R}\left(-\Omega,\mb{p}\right)\right]=\mathrm{Re}\left[t_{\R}\left(\Omega,\mb{p}\right)\right]$, then as discussed in Sec.~\ref{sec:Previous_Theory},
the AL contribution to the Hall conductivity vanishes~\cite{VarlamovBook}.
Note that this is an exact statement, regardless of the specific form the fluctuation propagator takes; in all fluctuation theories,
particle-hole asymmetry is required to obtain a non-vanishing AL Hall conductivity.
Finally, we remark that an alternative approach~\cite{Varlamov_Livanov_1991} investigated including particle-hole asymmetry not in the propagator itself but rather in the vertices.
As shown in Sec.~\ref{sec:Boson_LongCond}, the Ward identity relates the propagator and the vertices self-consistently.
Thus it is problematic if certain physics is retained in the vertex but not the propagator and vice-versa.

As in Sec.~\ref{sec:Boson_LongCond}, we now consider the small $\mu_{\pair}$ limit and approximate $\sinh(\beta\varpi/2)\approx\varpi/(2T)$.
This allows the frequency and momentum integrations in Eq.~(\ref{eq:sigmaxy_BosonKubo}) to be performed analytically, which for $d=2$ yields Eq.~(\ref{eq:sigmaxy_Boson}) and for $d=3$ the result is
\begin{equation}
\label{eq:sigmaxy_AL3d}
\frac{\sigma_{xy}^{b}}{B} = -\frac{\kappa(\kappa^{2}+\Gamma^{2})}{\Gamma^{2}}\frac{k_{B}T\left(e^{*}\right)^{3}}{96\pi \hbar c}\frac{1}{\sqrt{2M_{\pair}|\mu_{\pair}|^{3}}}.
\end{equation}
The constants $c, \hbar$, and $k_{B}$ have been restored to ensure Eq.~(\ref{eq:sigmaxy_AL3d}) has units of $s^{-1}$.
The bosonic contribution to the Hall conductivity is proportional to the signs of charge ($e^{*}$) and particle-hole asymmetry ($\kappa$).

\section{Fermionic electromagnetic response}
\label{sec:Fermion_Transport}
\subsection{Fermionic longitudinal conductivity}
\label{subsec:Fermion_LongCond}

\begin{figure}[h]
\includegraphics[width=.9\linewidth,clip]{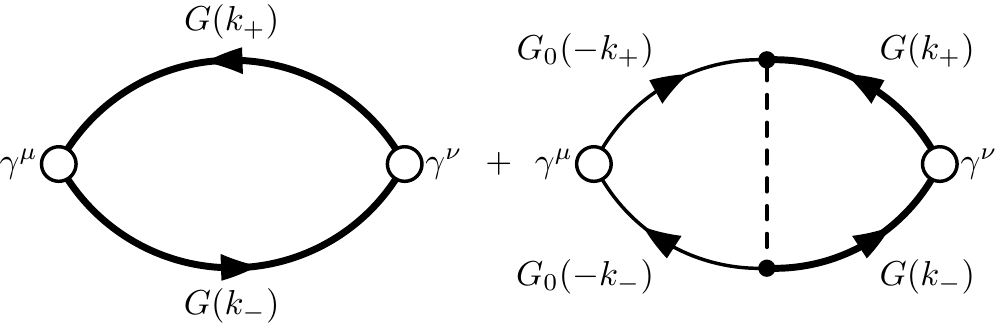}
\caption{Feynman diagram for the fermionic two-particle function.}
\label{fig:Fermion_xx}
\end{figure}

The fermionic contribution to the two-particle EM response is quite generally associated with DOS- and MT-type diagrams~\cite{Boyack_2018},
which in the presence of pseudogap effects are non-divergent.
For the conventional fluctuation theory the electrical conductivity of these diagrams has been well studied in the ultraclean~\cite{Varlamov_1991,Varlamov_Livanov_Savona_2000},
clean~\cite{Varlamov_Buzdin_1993,Varlamov_Livanov_Savona_2000},  and dirty~\cite{Varlamov_Buzdin_1993,VarlamovBook} cases.
However, in the strong-pairing fluctuation theory, where the $t$-matrix is given by Eq.~(\ref{eq:Tmat}), the presence of the normal-state pseudogap
makes the explicit calculations prohibitively difficult and so suitable approximations must be made for theoretical tractability.
We thus proceed on the basis of the well-studied fermionic quasiparticle picture~\cite{Tremblay_2017} and note that the
self energy $\Sigma(k)=\sum_{p}t(p)G_{0}(p-k)\varphi_{\mb{k}-\mb{p}/2}^2$  can be reasonably approximated~\cite{Maly_1999,Norman_2007} to be of the form:
$\Sigma(k)\approx-\left|\Delta\varphi_{\mb{k}}\right|^{2}G_{0}(-k)$, where $\left|\Delta\right|^{2}=-\sum_{p}t(p)$.
In this form the strong-pairing fluctuation theory has a physical interpretation associated with fermionic quasiparticles having a normal-state gap $\Delta$, while the bosonic fluctuations have been disregarded.
We have verified that the vertex corrections due to the form-factor contribution that arises for a $d$-wave pairing gap~\cite{Kosztin_2000} can be neglected in which case the fermionic two-particle correlation function is given by~\cite{Wulin_2012}:
\begin{align}
\label{eq:Fermion_Response}
P^{xx}_{f}(q) & =2e^{2}\sum_{k}\gamma^{x}(k_{+},k_{-})[G(k_{+})G(k_{-})\nonumber \\
 & \quad-F^{*}(k_{+})F(k_{-})]\gamma^{x}(k_{-},k_{+}).
\end{align}
The bare EM vertex is $\gamma^{x}(k_{+},k_{-})=\p\xi_{\mb{k}}/\p k^{x}=v_{x}$, where $\xi_{\mb{k}}$ is the single-particle bandstructure.
The diagram for the fermionic two-particle response is shown in Fig.~\ref{fig:Fermion_xx}.

The term $F$ has the exact functional form $F(k)=-\Delta\varphi_\mb{k}G_{0}(-k)G(k)$. While this is similar in appearance to an anomalous Green's function, it arises here due to a vertex correction associated with the pseudogap $\Delta$.
The first term in Eq.~(\ref{eq:Fermion_xx}) reflects a DOS-like interaction term and the second term is an MT-like diagram, now with the inclusion of the normal-state pseudogap.

The Matsubara frequency summation in Eq.~(\ref{eq:Fermion_Response}) is performed in the standard manner~\cite{MahanBook},
and after analytic continuation to real frequencies and then taking the limit $\mb{q}\rightarrow0$, the result obtained is:
\begin{widetext}
\begin{align}
\label{eq:Fermion_xx}
P^{xx}_{f}(\Omega,0) &= 2e^2\sum_{\mb{k}}v^{2}_{x}\int_{-\infty}^{\infty}\frac{d\omega}{2\pi}\tanh\left(\frac{\beta\omega}{2}\right)\biggl\{\Im\ G_{\R}(\omega,\mb{k})\left[G_{\R}(\omega+\Omega,\mb{k})+G_{\A}(\omega-\Omega,\mb{k})\right]
\nonumber\\&\quad - \Im\ F_{\R}(\omega,\mb{k})\left[F_{\R}(\omega+\Omega,\mb{k})+F_{\A}(\omega-\Omega,\mb{k})\right]\biggr\}.
\end{align}
\end{widetext}
After taking the limit $\Omega\rightarrow0$ in Eq.~(\ref{eq:Fermion_xx}) and then integrating by parts, the fermionic contribution to the longitudinal electrical conductivity is
\begin{align}
\label{eq:sigmaxx_FermionKubo}
\sigma^{f}_{xx} & =2e^{2}\sum_{\mb{k}}v^{2}_{x}\int_{-\infty}^{\infty}\frac{d\omega}{\pi}\left(-\frac{\p f(\omega)}{\p\omega}\right)\nonumber \\
 & \quad\times\left\{\left[\Im\ G_{\R}(\omega,\mb{k})\right]^{2}-\left[\Im\ F_{\R}(\omega,\mb{k})\right]^{2}\right\}.
\end{align}
The retarded propagators for the fermionic quasiparticles are
\begin{align}
\label{eq:Fermion_propagator}
G_{\R}(\omega,\mb{k}) & =\frac{u_{\mb{k}}^{2}}{\omega-E_{\mb{k}}+i\gamma}+\frac{v_{\mb{k}}^{2}}{\omega+E_{\mb{k}}+i\gamma},\\
F_{\R}(\omega,\mb{k}) & =-\frac{u_{\mb{k}}v_{\mb{k}}}{\omega-E_{\mb{k}}+i\gamma}+\frac{u_{\mb{k}}v_{\mb{k}}}{\omega+E_{\mb{k}}+i\gamma}.
\end{align}
The parameter $\gamma$ is related to the quasiparticle lifetime $\tau$ by $\gamma\equiv1/(2\tau)$. The coherence factors are $u_{\mb{k}}^{2}=\frac{1}{2}(1+\xi_{\mb{k}}/E_{\mb{k}})=1-v_{\mb{k}}^{2}$.
Inserting these propagators into Eq.~(\ref{eq:sigmaxy_FermionKubo}), and taking the limit where $\hbar\tau^{-1}\ll E_{F}$ to regularize the results~\cite{Voruganti_1992},
we obtain [as in Eq.~(\ref{eq:sigmaxx_Fermion})]:
\begin{equation}
\label{eq:sigmaxx_Fermion2}
\sigma^{f}_{xx}=2e^{2}\tau\sum_{\mb{k}}v^{2}_{x}\left(\frac{\xi_{\mb{k}}}{E_{\mb{k}}}\right)^{2}\left(-\frac{\p f(E_{\mb{k}})}{\p E_{\mb{k}}}\right).
\end{equation}
For the free-particle case, where $\Delta=0$ and $E_{\mb{k}}=|\xi_{\mb{k}}|=|k^{2}/(2m)-\mu|$, the electrical conductivity then reduces to the standard Drude expression $\sigma^{f}_{xx}=ne^{2}\tau/m$, as required.

\subsection{Fermionic Transverse conductivity }

The fermionic transverse conductivity is more complicated than that of the bosonic case due to the presence of a general bandstructure in contrast to an anisotropic but quadratic dispersion.
The general formalism for an arbitrary dispersion can be found in Ref.~\cite{Voruganti_1992} and we follow their methodology.
An additional complication is the incorporation of vertex corrections.
For the conventional fluctuation theory the Hall conductivity of the MT diagram has been studied in Ref.~\cite{Fukuyama_1971}.
In the strong-pairing fluctuation theory, however, the presence of the normal-state gap prevents an exact approach from being implemented.
A general investigation of incorporating vertex corrections was initiated in Ref.~\cite{Tremblay_2018}.
To make progress we follow Ref.~\cite{Tremblay_2017} and focus solely on the fermionic three-point function with three dressed Green's functions.
This response quite generally includes DOS-like effects, albeit without the incorporation of vertex corrections, and it incorporates the dominant effects from the gapped quasiparticles.
The Kubo-formula for the fermionic Hall conductivity is then~\cite{Voruganti_1992,Tremblay_2017}:
\begin{align}
\label{eq:sigmaxy_FermionKubo}
\frac{\sigma_{xy}^{f}}{B} &= \frac{4e^{3}}{3c}\sum_{\mb{p}}\int_{-\infty}^{\infty}\frac{d\omega}{\pi}\left(v_{x}^{2}v_{yy}-v_{x}v_{y}v_{xy}\right)\nonumber\\
&\quad \times\left[\Im\ G_{\R}(\omega,\mb{p})\right]^{3}\left(-\frac{\partial f(\omega)}{\partial \omega}\right).
\end{align}
Here we have assumed symmetry in the $x$-$y$ plane.
To regularize the product of three Green's functions appearing above, in the limit $\hbar\tau^{-1}\ll E_{F}$, we again use the method of Ref.~\cite{Voruganti_1992}; see the text below their Eq.~(1.26).
After inserting the retarded propagator from Eq.~(\ref{eq:Fermion_propagator}) into Eq.~(\ref{eq:sigmaxy_FermionKubo}) we obtain Eq.~(\ref{eq:sigmaxy_Fermion}):
\begin{align}
\label{eq:sigmaxy_Fermion2}
\frac{\sigma^{f}_{xy}}{B} &= \frac{e^{3}\tau^{2}}{2c}\sum_{\mb{k}}\left(v_{x}^{2}v_{yy}-v_{x}v_{y}v_{xy}\right)\left(1+\frac{3\xi_{\mb{k}}^{2}}{E_{\mb{k}}^{2}}\right)\nonumber \\
& \quad\times\left(-\frac{\partial f(E_{\mb{k}})}{\partial E_{\mb{k}}}\right).
\end{align}
For the free-particle case the transverse electrical conductivity reduces to the standard Drude expression $\sigma^{f}_{xy}/B=ne^{3}\tau^2/(m^{2}c)$, as required.

\section{Numerical Results}
\label{sec:Numerics}
\subsection{Phase diagram}

\begin{figure}
\includegraphics[width=0.6\linewidth]{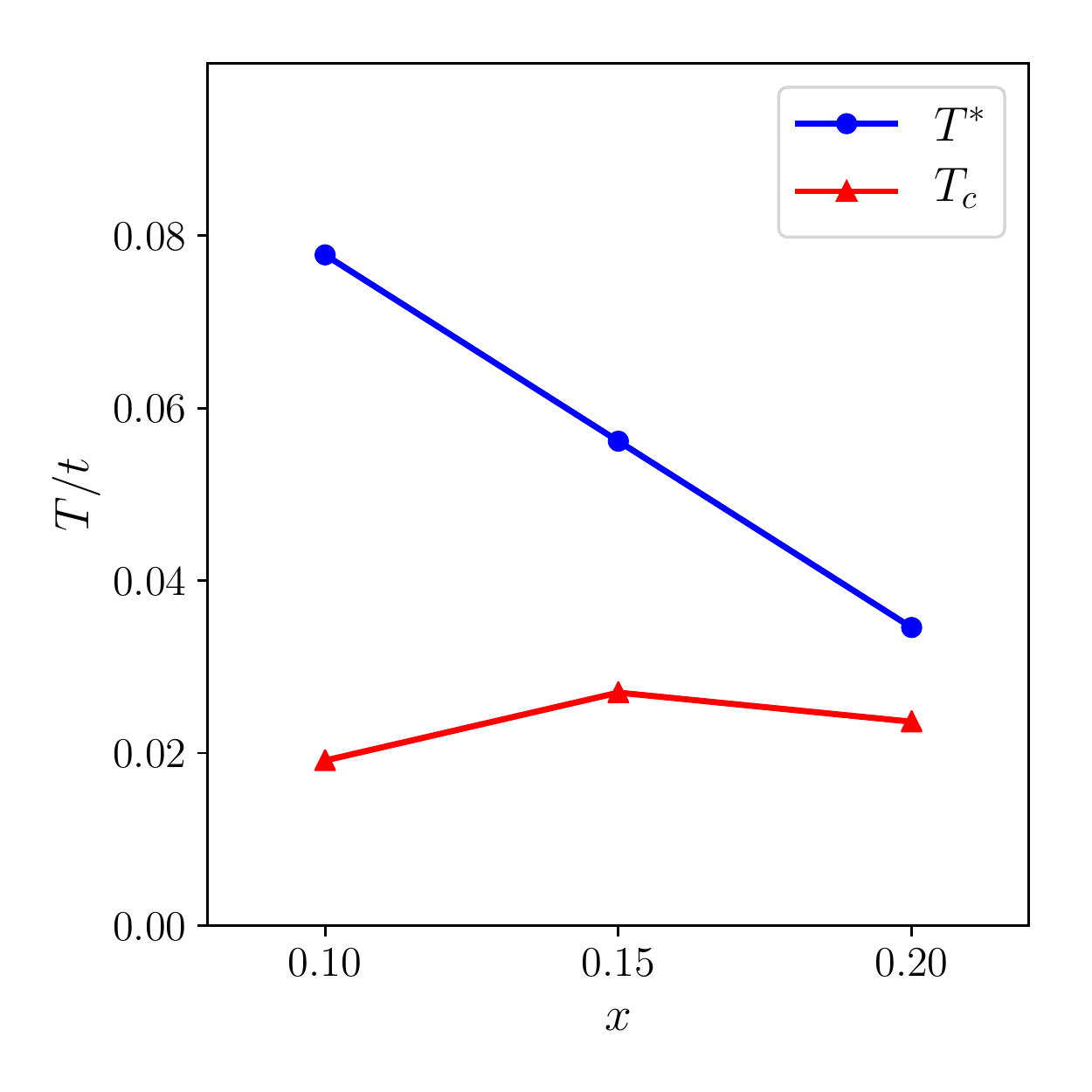}
\caption{The phase diagram for the strong-pairing fluctuation theory. The pairing-onset temperature is denoted $T^{*}$ whereas $T_{c}$ is the transition temperature, and here they are measured in terms of the hopping parameter $t$.
The hole concentration is labelled by $x$. The lines are a guide to the eye.}
\label{fig:phase_diagram}
\end{figure}

In this section we present the numerical analysis underlying the schematic illustration shown in Fig.~\ref{fig:schematic}.
Central to these results are the four equations given in Eqs.~(\ref{eq:sigmaxx_Boson}-\ref{eq:sigmaxy_Fermion}), which depend
on the fermionic dispersion and excitation gap $\Delta$ and on the bosonic parameters appearing in the pair propagator in Eq.~(\ref{eq:Tmat}).
Once these parameters are determined from the microscopic theory, a phase diagram can be computed for the temperature scales $T^*$ and $T_{c}$ at a few illustrative hole concentrations.
Throughout this section we use natural units: $e=\hbar=k_{B}=1$. The lattice constant $a$ and the hopping parameter $t$ are also set to 1.
When making comparisons with experimental numbers, one should appropriately restore the units, i.e., $a\approx 3$\AA \ and $t\approx 300 \mathrm{meV}$.

We do not have a microscopic theory to address the temperature dependence of the fermionic scattering time $\tau$, although its high-temperature limit is generally~\cite{TaoHu_2017}
taken as $\tau^{-1} = T$ for temperatures above $T^*$; see also Ref.~\cite{Taillefer_2018}.
For the purely-fermionic contribution to $R_{\H}$, however, this parameter cancels out. The calculations in this section should thus rather be viewed as qualitative.
The central goal is to arrive at a general understanding of the Hall coefficient over a broad range of temperatures, beyond that addressed in either the bosonic fluctuation literature or from the purely-fermionic perspective.

In support of our assertion that the high temperature upturn (with decreasing temperature) in $R_{\H}$ is associated with the fermionic contributions in Eqs.~(\ref{eq:sigmaxx_Fermion}-\ref{eq:sigmaxy_Fermion})
are widely observed~\cite{Hwang_1994,Vandermarel_1994,Xiang_2008} scaling observations which show how $R_{\H}$ varies with the pseudogap onset temperature $T^{*}$.
Additional experimental support for our results is provided from Ref.~\cite{Ando_Segawa_2002}, which presented a rather detailed set of plots for $\sigma_{xy}$
showing that it is positive and slightly increasing (with decreasing $T$) over a wide range of temperatures.
Related data for $\sigma_{xy}$ over a narrower temperature range were also presented in Ref.~\cite{Jin_Ott_1998}.
These experimental observations give credence to the claim that the bosonic contribution to Hall conductivity from the AL diagram has the same sign as the fermionic Hall response.
Moreover, away from the charge ordering regime, they give no indication for a divergence in $\sigma_{xy}$ (of either sign).

The microscopic model we adopt is the BCS Hamiltonian
\begin{equation}
H=\sum_{\mb{k},\sigma}\xi_{\mb{k}}c_{\mb{k}\sigma}^{\dagger}c_{\mb{k}\sigma}
-g\sum_{\mb{k},\mb{k'}}\varphi_{\mb{k}}\varphi_{\mb{k}'}c_{\mb{k}\uparrow}^{\dagger}c_{-\mb{k}\downarrow}^{\dagger}c_{\mb{k}'\downarrow}c_{-\mb{k'}\uparrow},
\end{equation}
where $g>0$ is an attractive interaction constant and $\varphi_\mb{k}=\cos k_x-\cos k_y$ is the $d$-wave form factor. The bare band dispersion is parameterized to be:
\begin{align}
\xi_{\mb{k}}&=2t\left(2-\cos k_{x}-\cos k_{y}\right)-2t'\left(1-\cos k_{x}\cos k_{y}\right) \nonumber\\
&\quad+2t_{z}\left(1-\cos k_{z}\right)-\mu,
\end{align}
with parameters $t=1$ and $t'=0.7$ consistent with a tight-binding fit to angle resolved photoemission (ARPES) measurements of YBCO~\cite{Tremblay_2017}.
The chemical potential is chosen such that the number of electrons per unit cell is $1-x$, where $x$ is the hole doping concentration.

The first numerical calculation we present is a phase diagram in Fig.~\ref{fig:phase_diagram}, obtained using our strong-pairing approach~\cite{Chen_Stajic_2005}.
The pseudogap temperature $T^{*}$ is the pairing-onset temperature and this crossover temperature can be estimated from the mean-field BCS gap equation with zero superconducting gap:
$g^{-1}= \sum_{\mb{k}}\frac{\varphi^{2}_{\mb{k}}}{E_{\mb{k}}}\tanh(\beta E_{\mb{k}}/2)$, where $E_{\mb{k}}=\sqrt{\xi^{2}_{\mb{k}}+|\Delta_{\mb{k}}|^{2}}$ is the quasiparticle dispersion, with $\Delta_\mb{k} = \Delta\varphi_\mb{k}$.
The superconducting transition temperature, $T_{c}$, is obtained by a self-consistent solution~\cite{Chen_Stajic_2005} associated with the condition $\mu_\text{pair}(T_{c})\equiv 0$.

The interaction strength $g(x)$ is chosen to capture the general features of the phase diagram; we fit it by taking $T^{*}(x)=0.432t\left(0.28-x\right)$.
In strictly two spatial dimensions we find $T_{c} = 0$, as is consistent with the Mermin-Wagner theorem.
The introduction of a small hopping constant along the $z$-axis, $t_z\ll t$, leads to finite, but low transition temperatures.
It is reasonable that they become lower with underdoping as the $c$-axis coupling is expected to be weaker.
These parameter choices are then sufficient to deduce the parameters relevant for transport calculations, i.e., $\mu_\text{pair}, M_\text{pair}, \Delta, T_{c}$, and $T^{*}$.
All of these are self-consistently calculated using the strong-fluctuation theory. For further details see Ref.~\cite{Chen_Stajic_2005}.

\subsection{Transport summary for optimal hole concentration}

\begin{figure}
\includegraphics[width=\linewidth]{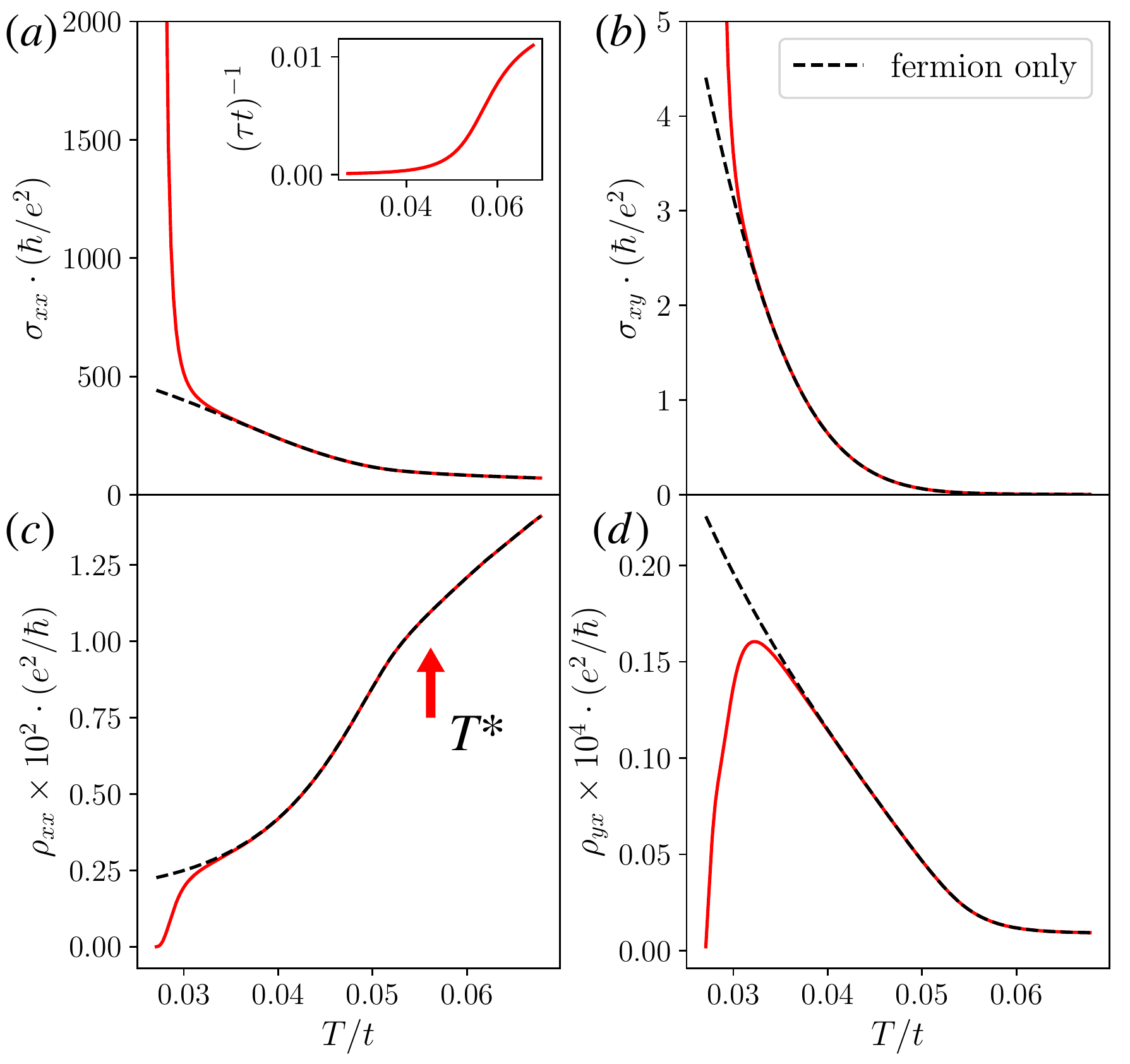}
\caption{Electrical transport properties $\sigma_{xx}$, $\sigma_{xy}$, $\rho_{xx}$, and $\rho_{yx}$, calculated for doping $x=0.15$.
The red curves combine both the fermionic and bosonic contributions to electrical transport, while the black dashed curves show only the fermionic contribution. The 2D conductivities are measured in units of $e^2/\hbar$.}
\label{fig:canonical}
\end{figure}

In Fig.~\ref{fig:canonical} we present the temperature dependence of transport quantities calculated for optimal hole doping $x=0.15$.
The effect of a pseudogap is to reduce the effective number of carriers and thereby to suppress $\sigma_{xx}$ (with decreasing $T$).
This gap effect is evidently countered by the temperature dependent lifetime,  as experiments show $\rho_{xx}$ is a monotonically decreasing function of decreasing temperature.
Here we phenomenologically introduce a temperature-dependent lifetime $\tau$ to arrive at a reasonable form for the longitudinal electrical  conductivity.
The specific form is given by $\tau^{-1}(T) =\frac{1}{4}T \exp(-\Delta(T)/T^*)$,  as depicted in the inset to Fig.~\ref{fig:canonical}.
The precise functional form for this lifetime is not important, as it cancels out when the purely-fermionic contribution to $R_{\H}$ dominates.
It is possible that the presence of Fermi arcs~\cite{Wulin_2011}--not captured in our analysis, would require a less dramatic temperature dependence in the lifetime.

Panels (a)-(d) in Fig.~\ref{fig:canonical} plot the longitudinal and transverse electrical conductivities, and
the longitudinal and transverse resistivities, respectively, as functions of temperature.
The dashed lines indicate the fermionic components. The conductivities are measured in units of $e^2/\hbar$.
A notable feature of the plot for $\rho_{xx}$ is that when compared with experimental values there is a difference of roughly a factor of 50.
This comes from the fact that the ARPES-derived bandstructure is associated with the full Luttinger volume ($1+x$ for holes)
whereas the plasma frequency which sets the scale for $\rho_{xx}$ suggests a reduced carrier number ($x$).
The change of Luttinger count with underdoping is not captured in our present study.
Importantly, this factor of 50 does not appear in $\rho_{yx}$ where there is semi-quantitative numerical agreement with experiment.

The two transverse response plots in Fig.~\ref{fig:canonical} are of the greatest interest.
While there has been extensive discussion about the sign of the bosonic (AL) contribution,
Ref.~\cite{Ando_Segawa_2002} presented a rather detailed set of plots for $\sigma_{xy}$ showing that it is positive and increasing (with decreasing $T$) over the wide range of temperatures.
This is consistent with the present theory and with the constraint we have emphasized that the sign of the bosonic contribution should be the same as that for the fermions; in this case it is positive.

Another observation from Fig.~\ref{fig:canonical} is that the fermionic and bosonic contributions to electrical transport are dominant in different temperature regions.
The upturn in $\rho_{yx}$ with decreasing $T$ reflects a pseudogap effect; as the excitation gap becomes larger,
the number of carriers decreases and the Hall coefficient necessarily increases.
Despite stronger-than-BCS fluctuations, coherent bosonic transport is still only dominant near $T_{c}$, as in the conventional fluctuation approach.
This bosonic contribution is responsible for the downturn in $\rho_{yx}$. In the linear response regime $\rho_{yx}$ is given by [Eq.~(\ref{eq:rhoyx})] $\rho_{yx}=\sigma_{xy}/\sigma^{2}_{xx}$,
and the large longitudinal electrical conductivity, in comparison to the transverse, is what causes this suppression in $\rho_{yx}$.
Notably, if we extend these expressions beyond the physical range of Eq.~(\ref{eq:EnergyScale}), both tend to diverge leading to a delicate competition which is regularized by Eq.~(\ref{eq:sigmaxx_alpha}).
This is consistent with experiment, which shows that pairing fluctuations appear only in the vicinity of the condensation temperature; they are uncorrelated with $T^{*}$.

\subsection{Doping and magnetic field effects on the Hall coefficient}

\begin{figure}
\includegraphics[width=\linewidth]{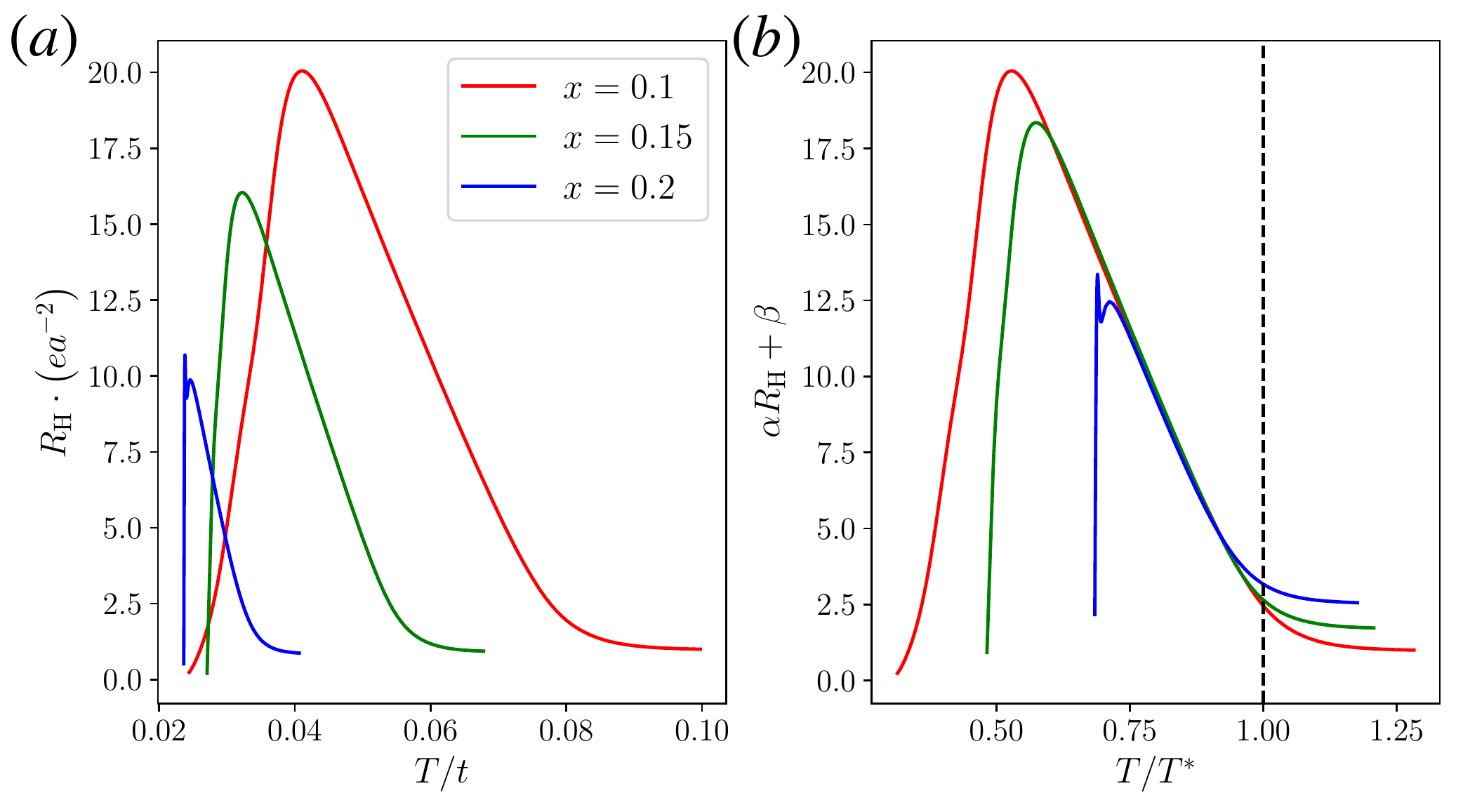}
\caption{(a) Temperature dependence of the Hall coefficient $R_{\H}$ at three doping levels $x=0.1$ (underdoped), $x=0.15$ (optimal), and $x=0.2$ (overdoped).
(b) Scaling of $R_{\H}$ with $T/T^*$. Here $\alpha$ and $\beta$ are scaling parameters to align the curves together, with $\alpha=1,1.1,1.2$, and $\beta=0,0.7,1.6$ from underdoped to overdoped.}
 \label{doping_scaling}
\end{figure}

In Fig.~\ref{doping_scaling}(a) the temperature-dependent Hall coefficient is presented for three representative doping levels.  All three curves share some common features.
As temperature is decreased $R_{\H}$ has an initial rise starting at the onset of the pseudogap at temperature $T^{*}$, followed by a rapid downturn in the vicinity of $T_{c}$.
The peak height increases with underdoping as the pseudogap is larger there and hence there are fewer carriers.
The small kink in the $x=0.2$ curve signals the onset of the contribution from $\sigma_{xy}^b$ which reflects a mismatch with the fermionic background.
In Fig.~\ref{doping_scaling}(b) it is shown how the fermionic (high-$T$) regime can be scaled onto a single ``backbone'' plot.
This scaling has received widespread interest~\cite{Hwang_1994,Vandermarel_1994,Xiang_2008} in the experimental community and serves to validate our finding here that
the behavior at the highest temperatures is associated with the fermionic degrees of freedom in the presence of an excitation gap with onset at $T^{*}$.
As might be expected, this scaling ceases at $T^{*}$, and above, when the Hall coefficients approach their different normal-state values.

\begin{figure}
\includegraphics[width=0.6\linewidth]{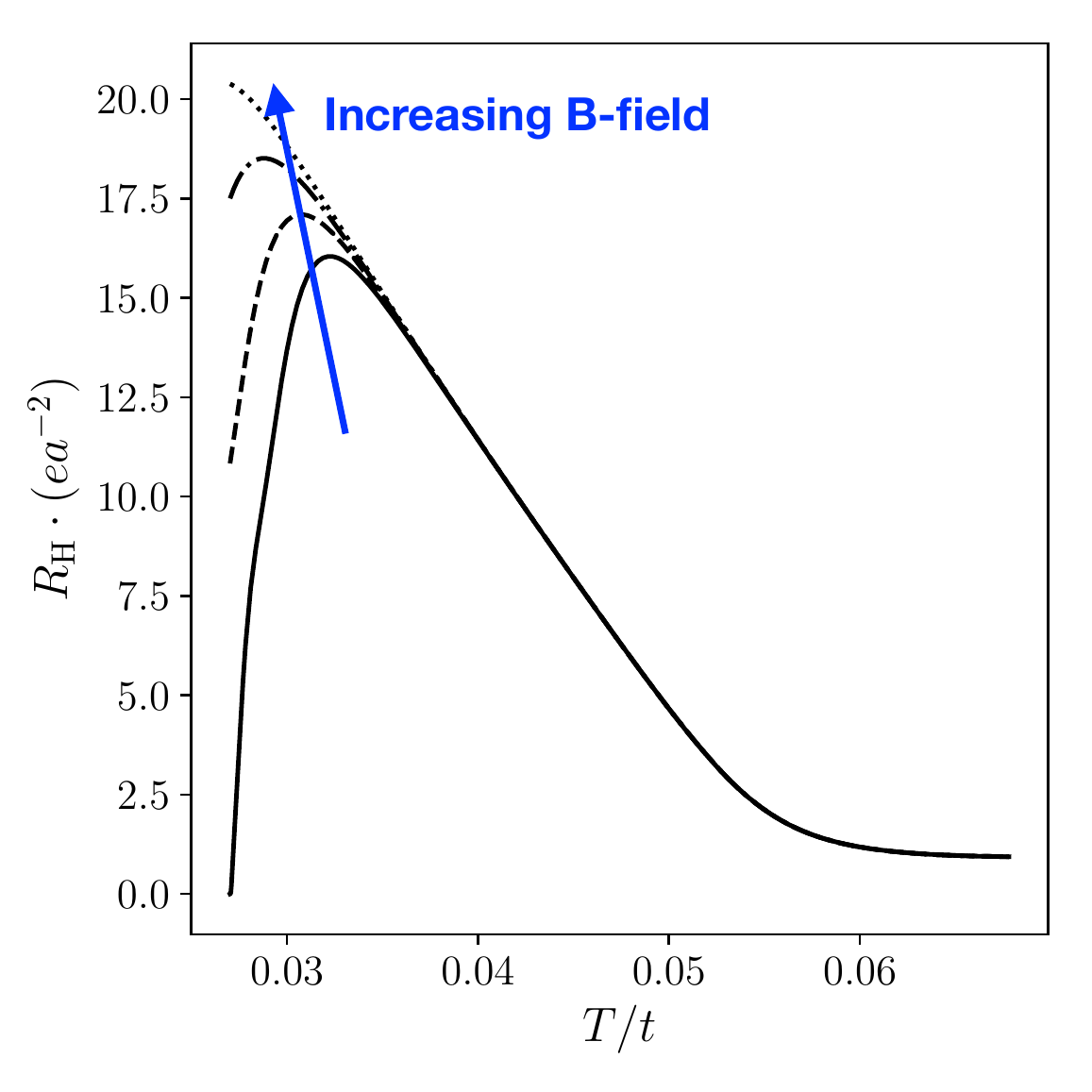}
\caption{Expected behavior of $R_{\H}$ in the presence of strong magnetic fields, modeled by a field-dependent $T_c$ and pair chemical potential $\mu_{\text{pair}}$, but with a pseudogap temperature $T^{*}$ that remains unaffected.
\label{fig:tc_scaling}}
\end{figure}

A schematic plot predicting the behavior of $R_{\H}$ at fixed stoichiometry as $T_{c}$ is increasingly depressed, while $T^{*}$ is relatively unaffected,  is presented in Fig.~~\ref{fig:tc_scaling}.
This situation is expected to pertain when the magnitude of an applied magnetic field is increased.
(We emphasize here that this plot is schematic since the present calculations are valid only in the linear response regime).
Nevertheless, the pairing onset temperature $T^{*}$ is relatively robust to variable magnetic fields, because this energy scale is large compared to typical magnetic field energies, even for fields as high as $50$T.
By contrast the coherence temperature $T_{c}$ is relatively more sensitive \cite{Shibauchi,Iyengar_2001}.
Thus the fermionic ``backbone", which depends on the pseudogap $\Delta$, might well be present even at the high fields recently investigated in Ref.~\cite{Taillefer_2016}.
We observe in Fig.~\ref{fig:tc_scaling} that the fermionic contribution is barely affected, while the downturn is suppressed to progressively lower temperatures with increasing field strength.

\section{Conclusions}
\label{sec:Conclusion}

In this paper we have studied the cuprate Hall coefficient.
Its non-monotonic behavior with temperature, the fact that it characterizes the sign of the charge carriers, and because
it establishes the degree of particle-hole asymmetry, make it an important quantity to study and one which allows microscopic properties of the cuprates to be addressed.
Transport properties such as longitudinal conductivity and diamagnetic susceptibility do not possess these features.
Thus, understanding the Hall coefficient provides deep insight into the cuprates and their mysterious pseudogap.

The literature has emphasized that there are two alternative approaches to addressing the cuprate Hall conductivity;
one based on the fermionic quasiparticle perspective~\cite{Hwang_1994,Vandermarel_1994,Xiang_2008,Taillefer_2016,Tremblay_2017,Mitscherling_2018}
and another focusing on bosonic Cooper-pair fluctuations~\cite{Rice_1991,Samoilov_1994,Lang_Heine_1994,Hwang_1994,Jin_Ott_1998,Konstantinovic_2000,Matthey_2001,Segawa_Ando_2004}.
In this paper it has been argued that the non-monotonicity in $R_{\H}$ can be understood only by including both fermionic and bosonic contributions.
Each type of excitation dominates in a different temperature regime and in the presence of a pseudogap these two are intimately related.
In particular, the pair propagator, which is at the heart of the bosonic fluctuation transport, must necessarily incorporate the non-vanishing excitation gap of the fermionic constituents.

We have also emphasized that there is no consensus in the literature on the relative sign between the bosonic fluctuation and fermionic quasiparticle contributions to $R_{\H}$.
A compelling argument adopted here is that the sign of the bosonic fluctuations is necessarily associated with that of the fermionic constituents and this sign is positive for the hole-doped cuprates with a hole-like Fermi surface.
As a result, there is no sign change in the normal-state Hall coefficient.
In this framework the experimentally observed decrease in $R_{\H}$ as $T_{c}$ is approached is interpreted as arising from gapless bosonic fluctuations with a longitudinal conductivity that is larger than the transverse conductivity.
In the weak magnetic field, linear response regime the Hall coefficient is $B R_{\H}\approx\sigma_{xy}/\sigma^{2}_{xx}$.
Experimental plots~\cite{Jin_Ott_1998,Ando_Segawa_2002} of $\sigma_{xy}$ and $R_{\H}$ indicate that it is the divergence in $\sigma_{xx}$ in the denominator which tends to dominate the
behavior of the Hall coefficient (and its rapid plummet) in the fluctuation regime.
Moreover, the sign (determined by that of $\sigma_{xy}$) near the transition and, also, well above, is consistent with hole-like quasiparticles, as we suggest here.

In summary, our paper has tackled the important problem of how to incorporate pseudogap effects into a fluctuation theory of Hall transport near $T_{c}$.
Furthermore, we have shown how to smoothly combine this with transport properties deriving from (pseudo)gapped fermionic quasi-particles which necessarily dominate at higher temperatures.
Future work through simultaneous Meissner and transport experiments in establishing $T_{c}$ and determining where exactly the sign change in $R_{\H}$ occurs relative to $T_{c}$ is needed;
this will aid in clarifying whether, as we have argued, the bosonic fluctuations have the same charge character as the fermionic constituents.

\acknowledgments

The authors wish to sincerely thank Christos Panagopoulos for sharing valuable insights into the experimental literature.
In addition we render our thanks to Alexey Galda, Suchitra Sebastian, Andrey Varlamov, and Tao Xiang for beneficial discussions.
XW and KL were supported by the University of Chicago Materials Research Science and Engineering Center, which is funded by the National Science Foundation under award number DMR-1420709.
RB was supported by the Theoretical Physics Institute at the University of Alberta.
QC was supported by NSF of China (Grant No.~11774309), and NSF of Zhejiang Province of China (Grant No.~LZ13A040001).

\appendix
\numberwithin{equation}{section}
\numberwithin{figure}{section}

\section{Gauge invariance of the bosonic three-particle EM response}
\label{sec:Gauge_Invariance}

\begin{widetext}
  In Sec.~\ref{sec:Boson_TranvCond}, the bosonic three-particle EM response function was proved to be gauge invariant in the small-$q$ limit.  In this appendix, a general proof of gauge invariance for arbitrary $q$ is provided.
  The bosonic three-particle EM response function is
\begin{align}
\label{eq:K_munualpha}
K^{\mu\nu\alpha}(q) &=
\left(e^{*}\right)^{3}\sum_{p}\biggl[\Lambda^{\mu}(i\varpi_{m}+i\Omega_{m},\mb{p}_{+};i\varpi_{m},\mb{p}_{-})\Lambda^{\nu}(i\varpi_{m},\mb{p}_{+};i\varpi_{m}+i\Omega_{m},\mb{p}_{+})\Lambda^{\alpha}(i\varpi_{m},\mb{p}_{-};i\varpi_{m},\mb{p}_{+})\nonumber \\
 & \quad\quad\quad\quad\times t(i\varpi_{m}+i\Omega_{m},\mb{p}_{+})t(i\varpi_{m},\mb{p}_{-})t(i\varpi_{m},\mb{p}_{+})\nonumber \\
 & \hspace{1.35cm}+\Lambda^{\mu}(i\varpi_{m},\mb{p}_{+};i\varpi_{m}-i\Omega_{m},\mb{p}_{-})\Lambda^{\nu}(i\varpi_{m}-i\Omega_{m},\mb{p}_{-};i\varpi_{m},\mb{p}_{-})\Lambda^{\alpha}(i\varpi_{m},\mb{p}_{-};i\varpi_{m},\mb{p}_{+})\nonumber \\
 & \quad\quad\quad\quad\times t(i\varpi_{m}-i\Omega_{m},\mb{p}_{-})t(i\varpi_{m},\mb{p}_{+})t(i\varpi_{m},\mb{p}_{-})\biggr]\nonumber \\
 & \quad +\frac{\left(e^{*}\right)^{3}}{M_{\pair}}\delta^{\nu\alpha}\sum_{p}\Lambda^{\mu}\left(i\varpi_{m}+i\Omega_{m},\mb{p}_{+};i\varpi_{m},\mb{p}_{-}\right)t(i\varpi_{m}+i\Omega_{m},\mb{p}_{+})t(i\varpi_{m},\mb{p}_{-}).
\end{align}
Here, $\mu,\nu,\alpha\in\{x,y\}$ and the four-vector $q^{\mu}=(i\Omega_{m},\mb{q})$. The Ward-Takahashi identity (WTI) relating the bosonic vertex to the bosonic propagator is
$q_{\mu}\Lambda^{\mu}(i\varpi_{m}^{+},\mb{p}_{+};i\varpi_{m}^{-},\mb{p}_{-})=t^{-1}(i\varpi_{m}^{+},\mb{p}_{+})-t^{-1}(i\varpi_{m}^{-},\mb{p}_{-}).$
By using the WTI, the contraction of Eq.~(\ref{eq:K_munualpha}) with $\mb{q}_{\alpha}$ is
\begin{align}
\label{eq:K_munualpha2}
\mb{q}_{\alpha}K^{\mu\nu\alpha}(q) &\propto \sum_{p}\left[t(i\varpi_{m},\mb{p}_{-})-t(i\varpi_{m},\mb{p}_{+})\right]\biggl\{\nonumber\\
&\quad\quad\Lambda^{\mu}(i\varpi_{m}+i\Omega_{m},\mb{p}_{+};i\varpi_{m},\mb{p}_{-})\Lambda^{\nu}(i\varpi_{m},\mb{p}_{+};i\varpi_{m}+i\Omega_{m},\mb{p}_{+})t(i\varpi_{m}+i\Omega_{m},\mb{p}_{+})\nonumber \\
&\quad+\Lambda^{\mu}(i\varpi_{m},\mb{p}_{+};i\varpi_{m}-i\Omega_{m},\mb{p}_{-})\Lambda^{\nu}(i\varpi_{m}-i\Omega_{m},\mb{p}_{-};i\varpi_{m},\mb{p}_{-})t(i\varpi_{m}-i\Omega_{m},\mb{p}_{-})\biggr\}\nonumber \\
&\quad- \frac{\mb{q}^{\nu}}{M_{\pair}}\sum_{p}\Lambda^{\mu}\left(i\varpi_{m}+i\Omega_{m},\mb{p}_{+};i\varpi_{m},\mb{p}_{-}\right)t(i\varpi_{m}+i\Omega_{m},\mb{p}_{+})t(i\varpi_{m},\mb{p}_{-}).
\end{align}
The spatial components of the vertices are $\Lambda^{\nu}(i\varpi_{m},\mb{p}_{+};i\varpi_{m}+i\Omega_{m},\mb{p}_{+})\propto(\mb{p}+\mb{q}/2)^{\nu}/M_{\pair}$
and similarly $\Lambda^{\nu}(i\varpi_{m}-i\Omega_{m},\mb{p}_{-};i\varpi_{m},\mb{p}_{-})\propto(\mb{p}-\mb{q}/2)^{\nu}/M_{\pair}$.
Inserting this into Eq.~(\ref{eq:K_munualpha2}) then gives
\begin{align}
\mb{q}_{\alpha}K^{\mu\nu\alpha}(q) &\propto \sum_{p}\left[t(i\varpi_{m},\mb{p}_{-})-t(i\varpi_{m},\mb{p}_{+})\right]\biggl\{\nonumber\\
&\quad\quad\Lambda^{\mu}(i\varpi_{m}+i\Omega_{m},\mb{p}_{+};i\varpi_{m},\mb{p}_{-})(\mb{p}+\mb{q}/2)^{\nu}t(i\varpi_{m}+i\Omega_{m},\mb{p}_{+})\nonumber \\
&\quad+\Lambda^{\mu}(i\varpi_{m},\mb{p}_{+};i\varpi_{m}-i\Omega_{m},\mb{p}_{-})(\mb{p}-\mb{q}/2)^{\nu}
 t(i\varpi_{m}-i\Omega_{m},\mb{p}_{-})\biggr\}\nonumber \\
 & \quad-\mb{q}^{\nu}\sum_{p}\Lambda^{\mu}\left(i\varpi_{m}+i\Omega_{m},\mb{p}_{+};i\varpi_{m},\mb{p}_{-}\right)t(i\varpi_{m}+i\Omega_{m},\mb{p}_{+})t(i\varpi_{m},\mb{p}_{-}).
\end{align}
In the third line let $\varpi_{m}\rightarrow\varpi_{m}+\Omega_{m}$ and then simplify the resulting equation to obtain
\begin{align}
\mb{q}_{\alpha}K^{\mu\nu\alpha}(q) &\propto \sum_{p}\Lambda^{\mu}(i\varpi_{m}+i\Omega_{m},\mb{p}_{+};i\varpi_{m},\mb{p}_{-})\biggl\{ t(i\varpi_{m}+i\Omega_{m},\mb{p}_{+})t(i\varpi_{m},\mb{p}_{-})
\left[(\mb{p}+\mb{q}/2)^{\nu}-(\mb{p}-\mb{q}/2)^{\nu}\right]\nonumber \\
 & \quad-(\mb{p}+\mb{q}/2)^{\nu}t(i\varpi_{m},\mb{p}_{+})t(i\varpi_{m}+i\Omega_{m},\mb{p}_{+})+(\mb{p}-\mb{q}/2)^{\nu}t(i\varpi_{m},\mb{p}_{-})t(i\varpi_{m}+i\Omega_{m},\mb{p}_{-})\biggr\}\nonumber \\
 & \quad-\mb{q}^{\nu}\sum_{p}\Lambda^{\mu}\left(i\varpi_{m}+i\Omega_{m},\mb{p}_{+};i\varpi_{m},\mb{p}_{-}\right)t(i\varpi_{m}+i\Omega_{m},\mb{p}_{+})t(i\varpi_{m},\mb{p}_{-}),\nonumber\\
 & =\sum_{p}\Lambda^{\mu}(i\varpi_{m}+i\Omega_{m},\mb{p}_{+};i\varpi_{m},\mb{p}_{-})\biggl\{(\mb{p}-\mb{q}/2)^{\nu}t(i\varpi_{m},\mb{p}_{-})t(i\varpi_{m}+i\Omega_{m},\mb{p}_{-})\nonumber \\
 & \quad-(\mb{p}+\mb{q}/2)^{\nu}t(i\varpi_{m},\mb{p}_{+})t(i\varpi_{m}+i\Omega_{m},\mb{p}_{+})\biggr\}.
 \end{align}
In the first term substitute $\mb{p}\rightarrow\mb{p}_{+}$ and in the second term substitute $\mb{p}\rightarrow\mb{p}_{-}$; the contraction of the response function is then
 \begin{equation}
\mb{q}_{\alpha}K^{\mu\nu\alpha}(q) \propto\sum_{p}t(i\varpi_{m},\mb{p})t(i\varpi_{m}+i\Omega_{m},\mb{p})\left((\mb{p}+\mb{q}/2)^{\mu}\mb{p}^{\nu}-(\mb{p}-\mb{q}/2)^{\mu}\mb{p}^{\nu}\right)= 0.
\end{equation}
In the last step let $\mb{p}\rightarrow-\mb{p}$ and use the fact the
fluctuation propagator depends on $\mb{p}^{2}$.  Therefore, the
bosonic three-particle EM response function is gauge-invariant.

\section{Hall conductivity calculations}
\label{sec:ALHall_Derivation}

In this excursus, the derivation of Eq.~(\ref{eq:AL_xy}) in the main text is presented. First consider the correlation function $K_{JJJ}$ presented in Eq.~(\ref{eq:KJJJ}) of the main text.
The first step is to perform the Matsubara frequency summation~\cite{VarlamovBook} and then perform the analytic continuation to real frequencies: $i\Omega_{m}\rightarrow\Omega+i0^{+}$.
The result that is obtained after this procedure is
\begin{align}
\label{eq:KJJJ_response}
K^{\mu\nu\alpha}_{JJJ}(\Omega,\mb{q}) &= \left(Ze^{*}\right)^{3}\sum_{\mb{p}}\left(\frac{\mb{p}^{\mu}}{M_{\pair}}\frac{\mb{p}^{\alpha}}{M_{\pair}}\right)\int_{-\infty}^{\infty}\frac{d\varpi}{2\pi}\coth\left(\frac{\beta\varpi}{2}\right)\nonumber \\
 & \quad\times\biggl\{\frac{\mb{p}_{+}^{\nu}}{M_{\pair}}\biggl[t_{\R}(\varpi+\Omega,\mb{p}_{+})\Im\left[t_{\R}(\varpi,\mb{p}_{-})t_{\R}(\varpi,\mb{p}_{+})\right]+t_{\A}(\varpi-\Omega,\mb{p}_{+})t_{\A}(\varpi-\Omega,\mb{p}_{-})\Im\left[t_{\R}(\varpi,\mb{p}_{+})\right]\biggr]\nonumber \\
 & \quad+\frac{\mb{p}_{-}^{\nu}}{M_{\pair}}\biggl[t_{\A}(\varpi-\Omega,\mb{p}_{-})\Im\left[t_{\R}(\varpi,\mb{p}_{+})t_{\R}(\varpi,\mb{p}_{-})\right]+t_{\R}(\varpi+\Omega,\mb{p}_{-})t_{\R}(\varpi+\Omega,\mb{p}_{+})\Im\left[t_{\R}(\varpi,\mb{p}_{-})\right]\biggr]\biggr\}.
\end{align}
Taking the limits $\mb{q}\rightarrow0$ followed by $\Omega\rightarrow0$ in Eq.~(\ref{eq:KJJJ_response}) then gives
\begin{align}
\label{eq:Re_KJJJ}
\mathrm{Re}K^{\mu\nu\alpha}_{JJJ}(\Omega,\mb{q}) &= \mb{q}^{\nu}\frac{\Omega\left(Ze^{*}\right)^{3}}{M_{\pair}}\sum_{\mb{p}}\left(\frac{\mb{p}^{\mu}}{M_{\pair}}\frac{\mb{p}^{\alpha}}{M_{\pair}}\right)\int_{-\infty}^{\infty}\frac{d\varpi}{2\pi}\ \coth\left(\frac{\beta\varpi}{2}\right)
\left[\Re(\p_{\varpi}t_{\R})\Im(t_{\R}^{2})-\Re(\p_{\varpi}t_{\R}^{2})\Im(t_{\R})\right],\nonumber \\
 & =\mb{q}^{\nu}\delta^{\mu\alpha}\frac{\Omega\left(Ze^{*}\right)^{3}}{dM_{\pair}}\sum_{\mb{p}}\left(\frac{p}{M_{\pair}}\right)^{2}\int_{-\infty}^{\infty}\frac{d\varpi}{2\pi}\coth\left(\frac{\beta\varpi}{2}\right)\left[\Re(\p_{\varpi}t_{\R})\Im(t_{\R}^{2})-\Re(\p_{\varpi}t_{\R}^{2})\Im(t_{\R})\right].
\end{align}

Now consider the correlation function $K_{J\rho}$ presented in Eq.~(\ref{eq:KJJJ}) of the main text.
After performing the Matsubara frequency summation and then the analytic continuation to real frequencies, the following result is obtained
\begin{equation}
\label{eq:KJrho_response}
K^{\mu\nu\alpha}_{J\rho}(\Omega,\mb{q}) = \frac{\left(Ze^{*}\right)^{3}}{M_{\pair}}\delta^{\nu\alpha}\sum_{\mb{p}}\frac{\mb{p}^{\mu}}{M_{\pair}}\int_{-\infty}^{\infty}\frac{d\varpi}{2\pi}\ \coth\left(\frac{\beta\varpi}{2}\right)
\biggl[t_{\R}(\varpi+\Omega,\mb{p}_{+})\Im\left[t_{\R}(\varpi,\mb{p}_{-})\right]+t_{\A}(\varpi-\Omega,\mb{p}_{-})\Im\left[t_{\R}(\varpi,\mb{p}_{+})\right]\biggr].
\end{equation}
Taking the limits $\mb{q}\rightarrow0$ followed by $\Omega\rightarrow0$ in Eq.~(\ref{eq:KJrho_response}) then gives
\begin{align}
\label{eq:Re_KJrho}
\mathrm{Re}K_{J\rho}^{\mu\nu\alpha}(\Omega,\mb{q}) &= \mb{q}^{\beta}\delta^{\nu\alpha}\frac{\Omega\left(Ze^{*}\right)^{3}}{M_{\pair}}\sum_{\mb{p}}\left(\frac{\mb{p}^{\mu}}{M_{\pair}}\frac{\mb{p}^{\beta}}{M_{\pair}}\right)\int_{-\infty}^{\infty}\frac{d\varpi}{2\pi}\ \coth\left(\frac{\beta\varpi}{2} \right)\left[\Re(\p_{\varpi}t_{\R}^{2})\Im(t_{\R})-\Re(\p_{\varpi}t_{\R})\Im(t_{\R}^{2})\right],\nonumber \\
 & =-\mb{q}^{\mu}\delta^{\nu\alpha}\frac{\Omega\left(Ze^{*}\right)^{3}}{dM_{\pair}}\sum_{\mb{p}}\left(\frac{p}{M_{\pair}}\right)^{2}\int_{-\infty}^{\infty}\frac{d\varpi}{2\pi}\coth\left(\frac{\beta\varpi}{2}\right)\left[\Re(\p_{\varpi}t_{\R})\Im(t_{\R}^{2})-\Re(\p_{\varpi}t_{\R}^{2})\Im(t_{\R})\right].
\end{align}
Adding Eq.~(\ref{eq:Re_KJJJ}) and Eq.~(\ref{eq:Re_KJrho}) then gives Eq.~(\ref{eq:AL_xy}) in the main text:
\begin{equation}
\mathrm{Re}K^{\mu\nu\alpha}\left(q\right)=\left(\mb{q}^{\nu}\delta^{\mu\alpha}-\mb{q}^{\mu}\delta^{\nu\alpha}\right)\frac{\Omega\left(Ze^{*}\right)^{3}}{dM_{\text{pair}}}\sum_{\mb{p}}\left(\frac{p}{M_{\text{pair}}}\right)^{2}
\int_{-\infty}^{\infty}\frac{dx}{2\pi}\coth\left(\frac{\beta\varpi}{2}\right)\left[\Re(\p_{\varpi}t_{\R})\Im(t_{\R}^{2})-\Re(\p_{\varpi}t_{\R}^{2})\Im(t_{\R})\right].
\end{equation}
\end{widetext}

\bibliography{Review}

\begin{thebibliography}{66}
\bibitem [{Com()}]{Comment3}%
  \BibitemOpen
  \href@noop {} {}\bibinfo {note} {It should be pointed out that this
  expression will not produce the BCS limit because $1/M_{\pair}$ increases
  with decreased pairing strength faster than $\Delta$ decreases. It is,
  however, useful to present this form in order to more readily transcribe the
  conventional fluctuation nomenclature.}\BibitemShut {Stop}%
\end{thebibliography}
\end{document}